\documentclass[aps,prd,amsmath,floats,floatfix,twocolumn,
superscriptaddress,nofootinbib,showpacs]{revtex4-1}

\usepackage[colorlinks, pdfborder={0 0 0}, plainpages=false]{hyperref}
\usepackage{graphicx}
\usepackage{xspace}
\usepackage[usenames,dvipsnames]{color}
\usepackage{amssymb}
\usepackage[dvipsnames]{xcolor}
\usepackage{placeins}
\usepackage[columnwise]{lineno} 
\usepackage{lineno}
\usepackage{amssymb}
\usepackage{amsfonts}
\usepackage{tensor}
\usepackage{bm}
\usepackage{units} 
\usepackage{amsopn, amstext, wasysym}
\usepackage{colonequals}
\usepackage{subfigure}
\usepackage{braket}
\usepackage{multirow}
\usepackage{xspace}
\usepackage[normalem]{ulem} 
\usepackage{ulem}

\newcommand{\Msun}{M_\odot}

\def\l({\left(}
\def\r){\right)}

\newcommand{\spec}{\texttt{SpEC}}

\newcommand{\NRSurL}{\texttt{NRSur7dq2L2}}
\newcommand{\NRSurHM}{\texttt{NRSur7dq2HM}}
\newcommand{\NRSur}{\texttt{NRSur7dq2}}
\newcommand{\Phenom}{\texttt{IMRPhenomPv2}}
\newcommand{\SEOB}{\texttt{SEOBNRv4}}
\newcommand{\SEOBNR}{\texttt{SEOBNR}}

\newcommand{\etal}{\textit{et~al}\@ifnextchar{\relax}{.\relax}{\ifx\@let@token.\else\ifx\@let@token~.\else.\@\xspace\fi\fi}}

\newcommand{\D}{\mathrm{d}}

\newcommand{\tht}{\vec{\theta}}
\newcommand{\tjn}{\theta_\mathrm{JN}}

\newcommand{\chieff}{\chi_\mathrm{eff}}
\newcommand{\chip}{\chi_\mathrm{p}}

\newcommand{\mchirp}{\mathcal{M}_c}

\newcommand{\tsyst}{$\overline{\delta\theta_\mathrm{syst}}$}
\newcommand{\tstat}{$\overline{\delta\theta_\mathrm{stat}}$}

\newcommand{\Caltech}{\affiliation{Theoretical Astrophysics,
    Walter Burke Institute for Theoretical Physics, MC 350-17,
    California Institute of Technology, Pasadena, CA 91125, USA}}
\newcommand{\Cornell}{\affiliation{Cornell Center for Astrophysics
    and Planetary Science, Cornell University, Ithaca, New York
    14853, USA}}
\newcommand{\AEI}{\affiliation{Albert Einstein Institute,
Am M\"uhlenberg, Golm, Germany}} %
\newcommand{\UMassD}{\affiliation{Department of Mathematics, University
    of Massachusetts, Dartmouth, MA 02747, USA}}
\newcommand{\JPL}{\affiliation{Jet Propulsion Laboratory,
    California Institute of Technology, Pasadena, CA 91109, USA}}

\begin{document}

\title{Constraining the parameters of GW150914 and GW170104 with
numerical relativity surrogates}

\author{Prayush Kumar}\email{prayush@astro.cornell.edu}\Cornell
\author{Jonathan Blackman}\Caltech
\author{Scott E. Field}\UMassD
\author{Mark Scheel}\Caltech
\author{Chad R. Galley}\JPL\Caltech
\author{Michael Boyle}\Cornell
\author{Lawrence E. Kidder}\Cornell
\author{Harald P. Pfeiffer}\AEI
\author{Bela Szilagyi}\Caltech\JPL
\author{Saul A. Teukolsky}\Cornell\Caltech

\date{\today}

\begin{abstract}
Gravitational-wave (GW) detectors have begun to observe coalescences of
heavy black hole binaries ($M\gtrsim 50M_\odot$) at a consistent pace
for the past few years. Accurate models of gravitational waveforms are
essential for unbiased and precise estimation of source parameters,
such as masses and spins of component black holes.
Recently developed surrogate models based on high-accuracy numerical
relativity (NR) simulations provide ideal models for constraining
physical parameters describing these heavy black hole merger events.
In this paper, we first demonstrate the viability of these multi-modal
surrogate models as reliable parameter estimation tools. We show that
within a fully Bayesian framework, {\it NR surrogates can help extract
additional information from GW observations that is inaccessible to
traditional models}. We demonstrate this by analyzing a set of synthetic
signals with NR surrogate templates and comparing against contemporary
approximate models.

We then consider the case of two of the earliest binary black holes
detected by the LIGO observatories, GW150914 and GW170104. We reanalyze their
data with the generically precessing NR-based surrogate model.
{\it We find that our refined analysis is able to extract information from
sub-dominant GW harmonics in data, and therefore better resolve the
degeneracy in measuring source luminosity distance and orbital inclination
for both events}. Our analysis estimates the sources of both events to be
$20-25\%$ further away than was previously estimated. Our analysis
also constrains their orbital orientations more tightly around face-on/off
configurations than before. Additionally, for GW150914 we constrain the
effective inspiral spin $\chieff$ more tightly around zero.
This work is one of the first to unambiguously extract sub-dominant
GW mode information from real events. It is also a first step toward
eliminating the approximations used in semi-analytic waveform models
from GW parameter estimation. It strongly motivates that NR surrogates
be extended to cover more of the binary black hole parameter space.
\end{abstract}


\maketitle

\section{Introduction}\label{s1:introduction}
General Relativity (GR) predicts that accelerated massive bodies emit energy in
the form of gravitational waves (GWs). In 2015, the first direct detection of
GWs coming from coalescing binary black holes (BBHs) was made by the LIGO
observatories~\cite{LIGOVirgo2016a}. Since then, many more GW signals from BBHs
have been observed by the LIGO and Virgo detectors~\cite{Abbott:2016nmj,
TheLIGOScientific:2016pea,Abbott:2017vtc,Abbott:2017oio}, ushering us into the
era of GW astronomy.
%
%
GW searches for BBH signals~\cite{Usman:2015kfa,Cannon:2011vi,
Privitera:2013xza}, the process of estimating their source properties~\cite{%
Veitch:2015}, as well as that of testing GR with them~\cite{%
TheLIGOScientific:2016src} rely heavily on the technique of
matched filtering, which tacitly assumes the availability of GW signal
models for BBHs.

For heavy black hole binaries (with masses $\gtrsim 50M_\odot$),
such as those that have dominated the event rates of LIGO-Virgo detectors so
far~\cite{TheLIGOScientific:2016pea,LIGOVirgo2018:GWTC_1}, a large fraction
of the observable signal consists of the last few tens of orbits prior to the
binary's merger. In this regime, the dynamical effects of GR are substantial,
making analytic treatment difficult.
Instead, numerical solutions of Einstein's equations~\cite{Pretorius2005a,
Pretorius2005b,Pretorius2005c,Campanelli2006a,Campanelli2006a,Baker2006a,
Baker-Campanelli-etal:2007} must be used.

The inspiral of a binary system of black holes along a quasi-circular
trajectory, and their subsequent merger and ringdown, is completely describable
by $8$ parameters - the masses of both holes, and their spin vectors.
Conventional parameter estimation (PE) algorithms search through this parameter
space to estimate parameters that best describe the signal embedded in
LIGO-Virgo data. In the process, they can take a large number
($\mathcal{O}(10^6)$) of steps, each requiring evaluation of a new waveform.
Although we have the technical capability to perform full numerical relativity
(NR) simulations over a good fraction of the multi-dimensional parameter space,
each simulation still takes a large amount of computing and human time.
Therefore, it has remained impractical to use NR simulations directly with
conventional PE methods for estimating physical parameters of BBHs.
There have been two possible alternatives that have been utilized
in the past: (a) using phenomenological waveform models containing free
parameters that are tuned to a (relatively) small number
of NR simulations~\cite{Buonanno:2009qa,Ajith-Babak-Chen-etal:2007}; and (b)
using grid-based parameter estimation methods~\cite{Pankow:2015cra} with
NR templates.
While
phenomenological waveform models \SEOBNR{} and \Phenom{} have been used
extensively in previous LIGO-Virgo publications~\cite{LIGOVirgo2016a,
LIGOVirgo2018:GWTC_1,LIGOScientific:2018jsj}, they still have many
shortcomings. On one hand, \Phenom{} uses a post-Newtonian theory
based waveform amplitude prescription, and captures BH spin effects
using only $2$ (of total $6$) spin degrees of freedom~\cite{Khan:2015jqa};
while on the other hand, only aligned-spin \SEOBNR{} models are
computationally inexpensive enough to be used in PE
analyses~\cite{TheLIGOScientific:2016wfe}.
Moreover, neither of them presently have precession dynamics near
merger fitted against NR simulations, and
have been shown to break down close to the parameter-space
boundary of their calibration domain~\cite{Boyle2007,Kumar:2016dhh}
or even within it~\cite{Kumar:2015tha}.
Grid-based PE methods (alternative (b)) have been recently applied to GW
observations~\cite{Abbott:2016apu,Lange:2017wki}; however, they have so far
only demonstrated the ability to constrain a subset of physical
parameters because of the sparse parameter space coverage of available NR
simulations.

A novel alternative arises from the development of surrogate
models~\cite{Field:2013cfa,Purrer:2014fza} for numerical relativity
waveforms~\cite{Blackman:2015pia}. Such data-driven models are constructed over
a training set of specially selected NR simulations. The waveforms from these
simulations are then ``interpolated" in parameter space. The resulting NR 
surrogate model is able to quickly generate new waveforms at arbitrary points
within the training region with the largest surrogate model errors typically
comparable to the largest errors in the numerical relativity simulations.
After a couple of simpler versions~\cite{Blackman:2015pia,Blackman:2017dfb},
Blackman et al~\cite{Blackman:2017pcm} published a surrogate model \NRSur{} for
generically spinning-precessing binaries. This \NRSur{} model was developed
using $744$ new simulations~\cite{Blackman:2015pia}
spanning a range of the $7$-dimensional space\footnote{All simulations can be
re-scaled to any point in the eight dimension of total mass $M$.} bounded in
mass ratio $q\leq 2$, and BH spin magnitudes $|\vec{\chi}_{1,2}|\leq 0.8$.
It provides all $\ell\leq 4$ waveform multipoles, and we use it in
two configurations: including all available modes (\NRSurHM{}) and including
only the dominant $\ell=2$ multipoles (\NRSurL{}). We emphasize that this is
the only model in literature that both includes $\ell>2$ GW modes and captures
unabbreviated BH spin dynamics through all $6$ degrees of freedom.
In this paper, we demonstrate the viability of using \NRSur{} in
the follow-up parameter estimation of GW signals coming from heavy BBHs such as
those that LIGO-Virgo have observed multiple times
already~\cite{LIGOVirgo2018:GWTC_1}. In particular, we use it to estimate
all physical parameters of the first two heavy BBH events GW150914 and GW170104,
significantly extending their past analyses~\cite{Abbott:2016apu,Lange:2018pyp,
Abbott:2017vtc}.

We first perform controlled tests by injecting $48$ synthetic GW signals
(details in Table~\ref{tab:injections005}) into
zero noise and inferring their source parameters with both \NRSur{} 
surrogate models. We compare these results against the \Phenom{} model, which
captures spin-orbit precession effects and has been used extensively in
recent LIGO-Virgo analyses~\cite{Schmidt2010,Schmidt:2014iyl,Khan:2015jqa}.
We vary the source parameters of injections as follows.
Mass ratio is varied from
$q=1.2-1.5$, source location between $500-1500$Mpc, orbital inclination
between close to face-on and edge-on configurations, and component spins are
chosen from four distinct configurations with magnitudes $0.4 - 0.65$. These
values are deliberately chosen to enhance spin-induced orbital precession.
We find that even \NRSurL{} can
noticeably improve on \Phenom{} when it comes to measuring masses and mass
ratios of binary sources out to $\sim 1$Gpc.
This can be seen from
Fig.~\ref{fig:m1_m2_all_limprior_injections005}. All other binary
parameters such as component spins and source location are recovered
consistently by both models.
We further find that the inclusion of higher order $\ell=\{3,4\}$ GW modes in
\NRSurHM{} allows us to measure luminosity distance and orbital inclination
more accurately for sources out to $\sim 1$Gpc. This improvement
is especially noticeable for edge-on configurations, which is expected since
higher order modes contribute relatively more when we observe the source
at a larger angle. For such sources, \NRSurHM{} also measures binary mass
ratios somewhat more precisely than \Phenom{} and
\NRSurL{}. Finally, we find that BH spins are measured broadly consistently
by all three models. For the closest sources (within $\sim 500$Mpc) we do
gain some additional information with \NRSurHM{} templates. This improvement
is, however, modest for the investigated cases and we expect it to be more
signficant for binary sources with higher
mass ratios for which sub-dominant modes carry a larger fraction of the total
signal power~\cite{PekowskyEtAl:2012,Healy:2013jza,Bustillo:2015qty}.
Since \NRSur{} is presently restricted to $1\leq q\leq 2$, our results provide
incentive for extending the domain of NR based surrogates to higher mass ratios.

Having established the performance of \NRSur{} surrogates within a fully
Bayesian parameter recovery framework, we next analyze the first-ever recorded
BBH merger event: GW150914. The primary improvement we note is in the estimation
of the binary's luminosity distance $d_L$ from Earth: with extra information
coming from sub-dominant modes, we are able to constrain $d_L$ close to
$\sim 530$Mpc, about $100$Mpc further away than all others models'
estimation. Simultaneously, the surrogate also constrains the source of GW150914
to be either face-on or face-off\footnote{We use ``face-on'' to mean that the
observer is close to the north pole of the binary ($\theta_{JN}$ close to $0$
degrees), and ``face-off'' to mean that the observer is close to the binary's
south pole ($\theta_{JN}$ close to $180$ degrees).} more strongly than other
models, disfavoring
edge-on configurations. Consistent with this, the \NRSur{} models estimate
the source's chirp/total mass to be marginally higher than what approximate
models measure. And finally, having complete $2$-body spin information encoded
in them, the \NRSur{} models constrain the effective-spin of GW150914 to be
closer to zero than other models (with the same sampling priors). These results
continue to hold when we compare them with the LVC analysis of this event~\cite{TheLIGOScientific:2016wfe}.

We also analyze the second heavy BBH event GW170104 recorded by LIGO
detectors in early 2017. As for GW150914, we find that the surrogate
constrains the luminosity distance to this event to be larger (by $10\%$)
than what approximate models that include only $\ell=2$ GW modes do.
Similarly, it also constrains the source to be closer to face-on/off than
edge-on more strongly than other models. The estimation of mass parameters
is consistent between \NRSur{} and semi-analytic models, with the former only
recovering the portion of mass ratio posterior with support in $1\leq q\leq 2$.
Finally, we find the estimation of spin parameters to be remarkably similar
between \NRSur{}, \Phenom{} and \SEOB{}, with little extra information
coming from the use of NR surrogates. These results are all consistent with the
the first analysis of this event by the LVC~\cite{Abbott:2017vtc}. A summary
in the form of median estimates and symmetric $90\%$ credible intervals for
inferred quantities is given in Table~\ref{tab:parameters}.

From the analyses of GW150914 and GW1701014, 
we learn that one consistently recovers additional information that helps break
the luminosity distance - inclination degeneracy for BBH events with NR
surrogate templates, allowing us to constrain GW source and orientation
location better. We also learn that, in some cases, one could constrain BBH
effective spins better with the NR surrogates since they contain
unabbreviated nonlinear GR information. However, spin measurements are
sensitive to the choice of sampling priors employed~\cite{Chatziioannou:2018wqx,
Lange:2018pyp}, and we defer an investigation of their effect on spin inferences
for both events to future work. Our results are encouraging and we
propose that \NRSurHM{} and future NR surrogate models be used as part of
standard GW event follow-ups. We also encourage the NR community to further
the development of surrogate models to higher-mass ratios, so that more BBH
sources can be studied with them. In order to enable further analysis by the
community, we provide full posterior samples from Bayesian parameter estimation
of LIGO/Virgo data for GW150914 and GW170104, with \NRSurHM{} and \Phenom{}.
These can be obtained from
\url{https://github.com/prayush/GW150914_GW170104_NRSur7dq2_Posteriors}.

The remainder of this paper is organized as follows. In
Section~\ref{s1:preliminaries} we describe the surrogate and approximate waveform
models used in this paper, as well as the details of our Bayesian parameter
estimation machinery. In Section~\ref{s1:injections} we present results from
studies involving parameter recovery from synthetic signals.
In Section~\ref{s1:gw150914} and~\ref{s1:gw170104} we present results of our
re-analysis of GW150914 and GW1701014 using the new NR surrogate model. And 
finally, in Section~\ref{s1:discussion} we summarize our findings and present
the future outlook for this research.

\section{Preliminaries}\label{s1:preliminaries}

\subsection{Numerical Relativity Surrogates}\label{s2:nr_surrogates}

A surrogate waveform model is one that takes a set of pre-computed waveforms
generated by an underlying model as input, and interpolates in parameter space
between these waveforms to produce waveforms for arbitrary parameter values.
The underlying waveform model can be analytic, phenomenological, or purely
numerical. Surrogates can often be evaluated in a fraction of the time that
takes for the underlying model to generate a waveform, and was in fact
originally proposed as a way to reduce the computational cost of
otherwise expensive waveform models when used with MCMC-based parameter
estimation algorithms applied to GW events~\cite{Field:2013cfa}.
With interpolation comes an additional source of modeling error, called the
{\it surrogate} error. In principle, this error can be arbitrarily reduced
by using a sufficiently large set of pre-computed waveforms to cover
the parameter space. In practice, when using NR waveforms the cost may become
prohibitive.

The \NRSur{} model of Ref.~\cite{Blackman:2017pcm} spans the $7$-dimensional
space of spin-precessing non-eccentric black hole binaries. It is built from
the results of $744$ NR simulations performed using the Spectral Einstein Code
\spec{}~\cite{SpECwebsite} and has already found several 
applications~\cite{Gerosa:2018qay,Talbot:2018sgr}. That it spans all
spin-precession degrees of freedom
comes at the cost of limiting its domain to comparable mass ratios $q =
m_1/m_2\leq 2$ and black hole spins with magnitudes $\leq 0.8$ of their extremal
values.
The choice of NR simulations used to train this surrogate was based on a
combination of methods including sparse grids~\cite{smolyak1963quadrature,
bungartz2004sparse} (as detailed in Appendix A of~\cite{Blackman:2017pcm}), a
template-metric-type stochastic sampler, and existing NR simulations. Taken
together, these choices maximized the coverage of the binary parameter space
with as few simulations as possible while simultaneously keeping the surrogate
error sufficiently small.
Instead of modeling waveform modes directly across the parameter space, the
strategy of \NRSur{} is to interpolate quantities that have as little structure
(such as oscillations) as possible. Ref.~\cite{Blackman:2017pcm} constructs
surrogate models for combinations of waveform modes in the {\it coorbital frame},
as well as for orbital phase and spin-related quantities that are required to 
transform these modes back to an inertial frame. They choose to parameterize
these fits using {\it instantaneous} spins and mass ratio, instead of
{\it initial} spins, as they find this choice improves the quality of fits.
Therefore, evaluation of \NRSur{} requires {first} obtaining the full
time-evolution of BH spins, orbital phase, and the unit quaternion that defines
the coprecessing frame~\cite{Boyle:2011gg,Schmidt2010,OShaughnessy2011}; and
{ subsequently} using these evolutions to
construct full inertial frame waveform modes from surrogate evaluation of
coorbital frame modes. We refer the reader to~\cite{Blackman:2017dfb,Blackman:2017pcm}
for further technical details and reasoning supporting various choices of
surrogate construction.
Finally, we note that \NRSur{} is limited in length to span the last $20$
binary orbits before merger. In practice, with a lower frequency cutoff of $20$
Hz, this restricts its use to binaries with total masses $M=m_1+m_2\gtrsim
50M_\odot$. In the remainder of the paper,
we will use \NRSurL{} to mean the surrogate with only $\ell=2$ waveform modes 
included, \NRSurHM{} for the surrogate with all its available $\ell=\{2,3,4\}$
modes, and \NRSur{} when discussing the surrogate model in general.

\subsection{Analytic Waveform Models}\label{s2:waveform_models}

In this paper, we will consider two waveform families: Effective-One-Body (EOB)
and phenomenological (IMRPhenom)~\cite{Buonanno99,Ajith-Babak-Chen-etal:2007}.
Both of these are semi-analytic models of the complete inspiral-merger-ringdown
for spinning BHs with non-eccentric orbits. For both models, we consider the
dominant
$(\ell,m)=(2,\pm 2)$ spin-weighted spherical harmonic waveform multipoles as only
these have been calibrated to NR simulations through merger.

{\bf Effective-One-Body:} 
The effective-one-body approach solves for the dynamics of the two-body problem
in nonlinear GR by mapping it to the dynamics of an {\it effective} test
particle of mass $\mu = m_1 m_2 / (m_1 + m_2)$ and spin 
$S^*(m_1, m_2,\vec{\chi}_1, \vec{\chi}_2)$
in a background spacetime that is described by a
parameterized deformation of the Kerr metric. Both $S^*$ and the background
deformation (to leading order) are chosen so that the geodesic followed
by the test particle reproduces the perturbative post-Newtonian (PN) dynamics of
the original two-body system~\cite{Blanchet:2013haa}. This conserved dynamics of
the test particle is described by the EOB Hamiltonian, which is also derived to
leading order using PN results. The radiative dynamics is introduced through a
flux of energy to emitted gravitational radiation, obtained by summing over all
PN-expanded waveform modes at future null infinity.
All of these model pieces are individually taken beyond known PN orders through
resummation and addition of phenomenological parameters that are subsequently
calibrated to ensure agreement of the inertial-frame waveform multipoles with NR
simulations. This allows the EOB prescription to be extended beyond the
slow-motion regime where PN results are valid, all the way up till the two BHs
merge. 
After merger, the ringdown waveform is constructed as a linear superposition of
the first eight quasi-normal modes (QNMs) of the Kerr BH formed at
merger~\cite{Kokkotas1999}. This ringdown waveform is suitably matched with the
inspiral-merger portion by enforcing
continuity of waveform modes and their first time-derivatives.

We use the most recent \SEOB{} model~\cite{Bohe:2016gbl} (available within the
LIGO Algorithms Library (LAL)~\cite{LAL}) in this study. This model describes
BBHs with component spins parallel to the orbital angular momentum (i.e.,
non-precessing binaries), on non-eccentric orbits, and was calibrated to $141$
NR simulations. We refer the reader to~\cite{Bohe:2016gbl} and references
therein for a comprehensive description of the model. In the interest of 
minimizing computational cost, we use the reduced-order model for \SEOB{}
that was also introduced in~\cite{Bohe:2016gbl}. We, however, are unable
to use the precessing EOB model of~\cite{Pan:2013rra} in this study due to its
high computational cost.

{\bf Phenomenological model:}
\Phenom{} is a phenomenological model constructed in the frequency domain that
describes GWs emitted by non-eccentric spinning-precessing binaries during
their inspiral-merger and ringdown phases~\cite{Schmidt2010,Schmidt:2014iyl,
Khan:2015jqa}.
It relies on the approximation that a generic precessing-binary inspiral
waveform
can be obtained by rotating the waveform for an equivalent spin-aligned
system in its quadrupole-aligned frame to the inertial frame using
time-dependent rotors (c.f. PN theory)~\cite{Schmidt2010,
Schmidt:2014iyl}.
In the quadrupole-aligned frame, leading order $(\ell,m)=(2,\pm 2)$ modes of
the waveform are constructed using the non-precessing IMRPhenomD
model~\cite{Khan:2015jqa}. The IMRPhenomD model has a closed form in frequency
domain, constructed piecewise in three portions: (i) early inspiral: where
both mode amplitude and phasing are given by extensions of PN-theory
results; (ii)-(iii) late inspiral and ringdown: where phenomenological ansatzes
are taken for waveform amplitude and phasing, and calibrated to enforce
high-precision agreement with NR simulations from various numerical relativity
groups. 
Note that \texttt{IMRPhenomD} captures BH spin effects on binary inspirals
using the effective-spin combination $\chieff:= (m_1\chi_{1z}+
m_2\chi_{2z})/M$, while \Phenom{} uses a precessing-spin parameter
$\chi_p$~\cite{Schmidt:2014iyl} to capture the precession of the
quadrupole-aligned frame with respect to inertial observers (instead
of using individual BH spin vectors).
Also
note that \Phenom{}({D}) belong to the unique class of models that are both
closed-form in the frequency domain {\it and} describe the complete
inspiral-merger-ringdown of spin-precessing binaries. These features are ideally
suited for GW searches and parameter estimation, which could require the
generation of a large number of waveform templates for each event.

\subsection{Parameter Estimation Methodology}\label{s2:bayesian_methods}

Let us denote the collection of measured parameters that describe a GW signal
received from a BBH merger event (including the binary's dynamical and
kinematic parameters, and other detector-related parameters\footnote{such as
those that describe instrument calibration uncertainty~\cite{Vitale:2011wu,
2013PhRvD..88h4044L}. For these, we take a conservative estimate of $10\%
/ 10^\circ$ uncertainty in amplitude / phase calibration for both LIGO
detectors~\cite{2013PhRvD..88h4044L}}) as $\tht$. The problem statement for
PE is to estimate the probability distribution $p(\tht)$ for the source binary.
Using Bayes' theorem, this posterior probability distribution $p(\tht)$ given
data $s$ from GW detectors containing the signal, and a model for GW signals
$H$ can be constructed as
\begin{equation}
 p(\tht|s,H) = \dfrac{p(s|\tht,H)\,p(\tht|H)}{p(s|H)},
\end{equation}
where: (i) $p(s|H)$ is the {\it prior} expectation of obtaining the new data
$s$, (ii) $p(\tht|H)$ is the expectation on parameters $\tht$ for astrophysical
sources {\it prior} to obtaining the new data, and (iii) $p(s|\tht,H)$ is the
likelihood of obtaining data $s$ ($=$ signal + noise) given $\tht$ describes
the signal embedded in it. Assuming the detector noise is stationary
colored-Gaussian with zero-mean, we can write
\begin{equation}\label{eq:likelihood}
 p(s|\tht,H) \propto \mathrm{exp}( - \frac{1}{2}\langle s-h_H(\tht)|s-h_H(\tht) \rangle ),
\end{equation}
where $h_H(\tht)$ is the signal waveform generated with the chosen GW model
$H$, and the noise-weighted inner product $\langle \cdot | \cdot\rangle$
between $a$ and $b$ is defined as
\begin{equation}\label{eq:inner_product}
 \langle a | b\rangle = 4 \Re\int_{f_l}^{f_{u}} \dfrac{a(f) b^*(f)}{S_n(f)} \D f,
\end{equation}
with $S_n(f)$ representing the one-sided power spectral density (PSD) of
detector noise
for LIGO. In this study, we use $f_u$ as the Nyquist frequency corresponding
to a sampling rate of $4096$Hz. We use the zero-detuning high power design
sensitivity curve for Advanced LIGO~\cite{aLIGO1,aLIGO2} when not using
detector data, and use $f_l=20$Hz as the lower frequency cutoff. For both events
(GW150914 and GW170104) we use LIGO data from its open science
center~\cite{Vallisneri:2014vxa}, and estimate detector PSD using $1024$
seconds of data around the signal concerned as described in~\cite{Veitch:2015}.

We compute $p(\tht|s,H)$ using the Bayesian inference package
LALInference~\cite{Veitch:2015} that is available as part of the LALSuite
software library~\cite{LAL}. LALInference has been extensively used
in past analyses published by the LIGO-Virgo Collaborations~\cite{LIGOVirgo2016a,
Abbott:2016nmj,TheLIGOScientific:2016pea,Abbott:2017vtc,Abbott:2017oio},
and uses the nested sampling algorithm~\cite{Skilling_NS} to estimate
source parameters from GW data. We refer the reader to~\cite{Veitch:2015}
for details of its implementation. As was its original purpose, nested sampling
already computes the integrated evidence $Z\equiv p(s|H)$ of the model $H$.
While unimportant to the parameter estimation problem, $Z$ is the key quantity
of interest for the purpose of model selection.


For all analyses in this article, we choose sampling priors $p(\tht |H)$
identical to those chosen in recent LIGO-Virgo results papers~\cite{%
TheLIGOScientific:2016wfe,Abbott:2017vtc}, i.e., both BH
masses and spin magnitudes are sampled uniformly over their respective ranges,
while spin directions are chosen uniformly over a $2-$sphere; source distance
and sky location are sampled uniformly in $3-$D spatial volume out to $2000$Mpc,
initial inclination angle is sampled uniformly from $[0,\pi]$, and the
remaining kinematic parameters are sampled uniformly over their respective
ranges. While these priors allow for a direct comparison of our results with
published LVC analyses~\cite{TheLIGOScientific:2016wfe,Abbott:2017vtc}, it has
been shown~\cite{Chatziioannou:2018wqx,Lange:2018pyp} that our (common)
choice of priors
downweights highly spinning binaries for which different choices of prior could
improve spin estimation. While neither of these work suggest that GW150914 or
GW170104 had large spins, we defer a rigorous study of the effect of priors on
the inference of BH spins for these events to future work.

We note from Sec.~\ref{s2:nr_surrogates} that \NRSur{} is limited to span
approximately $40$ GW cycles (of the $\ell=|m|=2$ modes) before merger. Therefore,
if the stochastic sampler of LALInference samples a point in binary parameter
space for which the complete waveform starting at $20$Hz is longer than $40$
cycles, the integrated likelihood (c.f. Eq.~\ref{eq:inner_product})  is
automatically reduced due to a reduction of the integration bandwidth to start
at the surrogate start frequency instead of $20$Hz.
We do not, however, {\it a priori} reject such a jump proposal.
Finally, we also note that waveform modes included in \NRSur{} templates that
have $m>2$ (such as the $(3,3), (4,4)$ modes) can start at frequencies above
$20$Hz. In order to mitigate the Gibbs phenomena brought on by the sudden start
of these higher-$m$ modes, we taper all templates at their start. However, some
of the information in these modes, contained in frequencies between $20$Hz and
their start, will be ignored in our analyses (as in previous analyses with
numerical simulations~\cite{Abbott:2016apu}). We expect though that these modes
contribute the most near merger and that the effect of missing lower frequencies
should be minimal~\cite{Abbott:2016apu}.
\setlength{\tabcolsep}{10pt}
\begin{table*}[]
\centering
\caption{\label{tab:injections005}
Parameters of injection set. Total $48$ injections.
Sources' sky location is chosen
from a uniform distribution over a $2-$sphere, while their polarization angle,
i.e. the third Euler angle required to rotate from the source to detector
frame, is chosen from a uniform distribution over $[0, \pi]$. Choices for these
three are held fixed for all injections. Total mass is held fixed at $60M_\odot$.
Their combined network SNR ranges from $\rho=13-83$ for the two-detector
Advanced LIGO network.
}
\begin{tabular}{|c|c|c|c|c|c|c|}
\hline\hline
Injection \# & $q\equiv m_1/m_2$ & $\vec{\chi}_1$ & $\vec{\chi}_2$ & $\tjn$ & D (Mpc) & Signal model \\ \hline
$0$ &  $1.2$ & $0.65\left(1, 1, 0\right) / \sqrt{2}$ & $-0.65\left(1, 1, 0\right)/ \sqrt{2}$ & $30^\circ$ & $1500$ & \NRSurHM{} \\
$1$ &  $1.2$ & $0.65\left(1, 1, 0\right) / \sqrt{2}$ & $\left(0, 0, 0\right)$ & $30^\circ$ & $1500$ & \NRSurHM{} \\
$2$ &  $1.2$ & $0.4\left(1, 1, 1\right)/ \sqrt{3}$ & $0.4\left(1, 1, 1\right)/ \sqrt{3}$ & $30^\circ$ & $1500$ & \NRSurHM{} \\
$3$ &  $1.2$ & $0.4\left(1, 1, 1\right)/ \sqrt{3}$ & $-0.4\left(1, 1, 1\right)/ \sqrt{3}$ & $30^\circ$ & $1500$ & \NRSurHM{} \\
$4$ &  $1.5$ & $0.65\left(1, 1, 0\right) / \sqrt{2}$ & $-0.65\left(1, 1, 0\right)/ \sqrt{2}$ & $30^\circ$ & $1500$ & \NRSurHM{} \\
$5$ &  $1.5$ & $0.65\left(1, 1, 0\right) / \sqrt{2}$ & $\left(0, 0, 0\right)$ & $30^\circ$ & $1500$ & \NRSurHM{} \\
$6$ &  $1.5$ & $0.4\left(1, 1, 1\right)/ \sqrt{3}$ & $0.4\left(1, 1, 1\right)/ \sqrt{3}$ & $30^\circ$ & $1500$ & \NRSurHM{} \\
$7$ &  $1.5$ & $0.4\left(1, 1, 1\right)/ \sqrt{3}$ & $-0.4\left(1, 1, 1\right)/ \sqrt{3}$ & $30^\circ$ & $1500$ & \NRSurHM{} \\
$8$ &  $1.2$ & $0.65\left(1, 1, 0\right) / \sqrt{2}$ & $-0.65\left(1, 1, 0\right)/ \sqrt{2}$ & $30^\circ$ & $1000$ & \NRSurHM{} \\
$9$ &  $1.2$ & $0.65\left(1, 1, 0\right) / \sqrt{2}$ & $\left(0, 0, 0\right)$ & $30^\circ$ & $1000$ & \NRSurHM{} \\
$10$ &  $1.2$ & $0.4\left(1, 1, 1\right)/ \sqrt{3}$ & $0.4\left(1, 1, 1\right)/ \sqrt{3}$ & $30^\circ$ & $1000$ & \NRSurHM{} \\
$11$ &  $1.2$ & $0.4\left(1, 1, 1\right)/ \sqrt{3}$ & $-0.4\left(1, 1, 1\right)/ \sqrt{3}$ & $30^\circ$ & $1000$ & \NRSurHM{} \\
$12$ &  $1.5$ & $0.65\left(1, 1, 0\right) / \sqrt{2}$ & $-0.65\left(1, 1, 0\right)/ \sqrt{2}$ & $30^\circ$ & $1000$ & \NRSurHM{} \\
$13$ &  $1.5$ & $0.65\left(1, 1, 0\right) / \sqrt{2}$ & $\left(0, 0, 0\right)$ & $30^\circ$ & $1000$ & \NRSurHM{} \\
$14$ &  $1.5$ & $0.4\left(1, 1, 1\right)/ \sqrt{3}$ & $0.4\left(1, 1, 1\right)/ \sqrt{3}$ & $30^\circ$ & $1000$ & \NRSurHM{} \\
$15$ &  $1.5$ & $0.4\left(1, 1, 1\right)/ \sqrt{3}$ & $-0.4\left(1, 1, 1\right)/ \sqrt{3}$ & $30^\circ$ & $1000$ & \NRSurHM{} \\
$16$ &  $1.2$ & $0.65\left(1, 1, 0\right) / \sqrt{2}$ & $-0.65\left(1, 1, 0\right)/ \sqrt{2}$ & $30^\circ$ & $500$ & \NRSurHM{} \\
$17$ &  $1.2$ & $0.65\left(1, 1, 0\right) / \sqrt{2}$ & $\left(0, 0, 0\right)$ & $30^\circ$ & $500$ & \NRSurHM{} \\
$18$ &  $1.2$ & $0.4\left(1, 1, 1\right)/ \sqrt{3}$ & $0.4\left(1, 1, 1\right)/ \sqrt{3}$ & $30^\circ$ & $500$ & \NRSurHM{} \\
$19$ &  $1.2$ & $0.4\left(1, 1, 1\right)/ \sqrt{3}$ & $-0.4\left(1, 1, 1\right)/ \sqrt{3}$ & $30^\circ$ & $500$ & \NRSurHM{} \\
$20$ &  $1.5$ & $0.65\left(1, 1, 0\right) / \sqrt{2}$ & $-0.65\left(1, 1, 0\right)/ \sqrt{2}$ & $30^\circ$ & $500$ & \NRSurHM{} \\
$21$ &  $1.5$ & $0.65\left(1, 1, 0\right) / \sqrt{2}$ & $\left(0, 0, 0\right)$ & $30^\circ$ & $500$ & \NRSurHM{} \\
$22$ &  $1.5$ & $0.4\left(1, 1, 1\right)/ \sqrt{3}$ & $0.4\left(1, 1, 1\right)/ \sqrt{3}$ & $30^\circ$ & $500$ & \NRSurHM{} \\
$23$ &  $1.5$ & $0.4\left(1, 1, 1\right)/ \sqrt{3}$ & $-0.4\left(1, 1, 1\right)/ \sqrt{3}$ & $30^\circ$ & $500$ & \NRSurHM{} \\
$24$ &  $1.2$ & $0.65\left(1, 1, 0\right) / \sqrt{2}$ & $-0.65\left(1, 1, 0\right)/ \sqrt{2}$ & $75^\circ$ & $1500$ & \NRSurHM{} \\
$25$ &  $1.2$ & $0.65\left(1, 1, 0\right) / \sqrt{2}$ & $\left(0, 0, 0\right)$ & $75^\circ$ & $1500$ & \NRSurHM{} \\
$26$ &  $1.2$ & $0.4\left(1, 1, 1\right)/ \sqrt{3}$ & $0.4\left(1, 1, 1\right)/ \sqrt{3}$ & $75^\circ$ & $1500$ & \NRSurHM{} \\
$27$ &  $1.2$ & $0.4\left(1, 1, 1\right)/ \sqrt{3}$ & $-0.4\left(1, 1, 1\right)/ \sqrt{3}$ & $75^\circ$ & $1500$ & \NRSurHM{} \\
$28$ &  $1.5$ & $0.65\left(1, 1, 0\right) / \sqrt{2}$ & $-0.65\left(1, 1, 0\right)/ \sqrt{2}$ & $75^\circ$ & $1500$ & \NRSurHM{} \\
$29$ &  $1.5$ & $0.65\left(1, 1, 0\right) / \sqrt{2}$ & $\left(0, 0, 0\right)$ & $75^\circ$ & $1500$ & \NRSurHM{} \\
$30$ &  $1.5$ & $0.4\left(1, 1, 1\right)/ \sqrt{3}$ & $0.4\left(1, 1, 1\right)/ \sqrt{3}$ & $75^\circ$ & $1500$ & \NRSurHM{} \\
$31$ &  $1.5$ & $0.4\left(1, 1, 1\right)/ \sqrt{3}$ & $-0.4\left(1, 1, 1\right)/ \sqrt{3}$ & $75^\circ$ & $1500$ & \NRSurHM{} \\
$32$ &  $1.2$ & $0.65\left(1, 1, 0\right) / \sqrt{2}$ & $-0.65\left(1, 1, 0\right)/ \sqrt{2}$ & $75^\circ$ & $1000$ & \NRSurHM{} \\
$33$ &  $1.2$ & $0.65\left(1, 1, 0\right) / \sqrt{2}$ & $\left(0, 0, 0\right)$ & $75^\circ$ & $1000$ & \NRSurHM{} \\
$34$ &  $1.2$ & $0.4\left(1, 1, 1\right)/ \sqrt{3}$ & $0.4\left(1, 1, 1\right)/ \sqrt{3}$ & $75^\circ$ & $1000$ & \NRSurHM{} \\
$35$ &  $1.2$ & $0.4\left(1, 1, 1\right)/ \sqrt{3}$ & $-0.4\left(1, 1, 1\right)/ \sqrt{3}$ & $75^\circ$ & $1000$ & \NRSurHM{} \\
$36$ &  $1.5$ & $0.65\left(1, 1, 0\right) / \sqrt{2}$ & $-0.65\left(1, 1, 0\right)/ \sqrt{2}$ & $75^\circ$ & $1000$ & \NRSurHM{} \\
$37$ &  $1.5$ & $0.65\left(1, 1, 0\right) / \sqrt{2}$ & $\left(0, 0, 0\right)$ & $75^\circ$ & $1000$ & \NRSurHM{} \\
$38$ &  $1.5$ & $0.4\left(1, 1, 1\right)/ \sqrt{3}$ & $0.4\left(1, 1, 1\right)/ \sqrt{3}$ & $75^\circ$ & $1000$ & \NRSurHM{} \\
$39$ &  $1.5$ & $0.4\left(1, 1, 1\right)/ \sqrt{3}$ & $-0.4\left(1, 1, 1\right)/ \sqrt{3}$ & $75^\circ$ & $1000$ & \NRSurHM{} \\
$40$ &  $1.2$ & $0.65\left(1, 1, 0\right) / \sqrt{2}$ & $-0.65\left(1, 1, 0\right)/ \sqrt{2}$ & $75^\circ$ & $500$ & \NRSurHM{} \\
$41$ &  $1.2$ & $0.65\left(1, 1, 0\right) / \sqrt{2}$ & $\left(0, 0, 0\right)$ & $75^\circ$ & $500$ & \NRSurHM{} \\
$42$ &  $1.2$ & $0.4\left(1, 1, 1\right)/ \sqrt{3}$ & $0.4\left(1, 1, 1\right)/ \sqrt{3}$ & $75^\circ$ & $500$ & \NRSurHM{} \\
$43$ &  $1.2$ & $0.4\left(1, 1, 1\right)/ \sqrt{3}$ & $-0.4\left(1, 1, 1\right)/ \sqrt{3}$ & $75^\circ$ & $500$ & \NRSurHM{} \\
$44$ &  $1.5$ & $0.65\left(1, 1, 0\right) / \sqrt{2}$ & $-0.65\left(1, 1, 0\right)/ \sqrt{2}$ & $75^\circ$ & $500$ & \NRSurHM{} \\
$45$ &  $1.5$ & $0.65\left(1, 1, 0\right) / \sqrt{2}$ & $\left(0, 0, 0\right)$ & $75^\circ$ & $500$ & \NRSurHM{} \\
$46$ &  $1.5$ & $0.4\left(1, 1, 1\right)/ \sqrt{3}$ & $0.4\left(1, 1, 1\right)/ \sqrt{3}$ & $75^\circ$ & $500$ & \NRSurHM{} \\
$47$ &  $1.5$ & $0.4\left(1, 1, 1\right)/ \sqrt{3}$ & $-0.4\left(1, 1, 1\right)/ \sqrt{3}$ & $75^\circ$ & $500$ & \NRSurHM{} \\
\hline
\end{tabular}
\end{table*}
\setlength{\tabcolsep}{0pt}

\section{Parameter Estimation of Synthetic GW signals}\label{s1:injections}
Since the \NRSur{} surrogate model have not been used to extract BBH
parameters from GW signals before, we start with controlled tests using
synthetic GW signals to establish the model's viability and benefits
for this purpose. We know that \NRSur{} produces NR-level accurate
templates~\cite{Blackman:2017pcm}, and includes $\ell\leq 4$ waveform
modes. Therefore there are two reasons why performing parameter estimation
with \NRSur{} templates may furnish more accurate results than any
waveform model, since approximate models can still struggle with modeling
the highly dynamical merger regime~\cite{Kumar:2016dhh}, and none of the
precessing ones used in recent LIGO-Virgo papers include $\ell>2$
modes~\cite{LIGOVirgo2016a,Abbott:2016nmj,TheLIGOScientific:2016pea,
Abbott:2017vtc,Abbott:2017oio,LIGOVirgo2018:GWTC_1}.
With injection tests we will study both reasons together and highlight
their distinct effects wherever manifest. We will also investigate
when information is lost due to artificially restricted domain of validity
of \NRSur{}.

We perform a total of $48$ injections in zero noise\footnote{``zero noise'' 
implies that data is composed of an injected signal plus zeros. Since detector
noise is assumed colored-Gaussian with zero mean, using zero noise with a
detector-noise weighted likelihood in Eq.~\ref{eq:likelihood} makes our analysis
equivalent to the average over an  {\it ensemble} of analyses which use actual
noise-realizations~\cite{Abbott:2016wiq}.} for both LIGO detectors,
and analyze the resulting coincident synthetic data with the \NRSurL{} model,
the \NRSurHM{} model, and the precessing \Phenom{} model
(with and without artificially restricted priors from \NRSur{}). We do not
include \SEOB{} in this section as it only models non-precessing sources,
and could not include the precessing EOB model~\cite{Pan:2013rra} because of
its high computational cost of generation~\cite{Abbott:2016izl}.
All synthetic signals are generated with the most accurate model available,
i.e., \NRSurHM{} including all $\ell\leq 4$ modes. 
We use the design noise curve for Advanced LIGO detectors~\cite{aLIGO1,aLIGO2}
while filtering, with a lower frequency cutoff of $20$Hz.

For the analyses to be relevant to heavy BH binaries, we fix the total mass
to $M=60M_\odot$ for all signals. All other parameters are chosen in the
following manner.
Mass ratio is chosen from two values $q=\{ 1.2, 1.5\}$. BH spins are sampled
from $4$ distinct configurations: (i) both spins with magnitude $0.65$ and
both initially parallel to orbital plane with $\hat{\chi}_1=-\hat{\chi}_2$;
(ii) spin on bigger BH with magnitude $0.65$ and parallel to orbital plane,
with $\vec{\chi}_2=0$; (iii) both spins with magnitude $0.4$ and mutually
parallel with $\hat{\chi}_1=\hat{\chi}_2=\frac{1}{\sqrt{3}}(1,1,1)$; and (iv)
both spins with magnitude $0.4$ and anti-parallel with $\hat{\chi}_1=
-\hat{\chi}_2=\frac{1}{\sqrt{3}}(1,1,1)$. These spin configurations are chosen
to enhance the effects of spin-induced orbital precession. The initial 
inclination of binary's total angular momentum with the detectors' line of sight
is chosen from two values, one close to nearly face-on with $\tjn = 30^\circ$ and
another close to edge-on with $\tjn = 75^\circ$. Each chosen source is then
placed at a luminosity distance $d_L=\{500,1000,1500\}$Mpc from the detectors.
Together these choices form a grid of $2\times 4\times 2\times 3 = 48$
injections. We list these parameter choices for injections in
Table~\ref{tab:injections005}.

\begin{figure}
 \centering
 \includegraphics[width=0.96\columnwidth,clip=true,trim={0 0 0 0}]{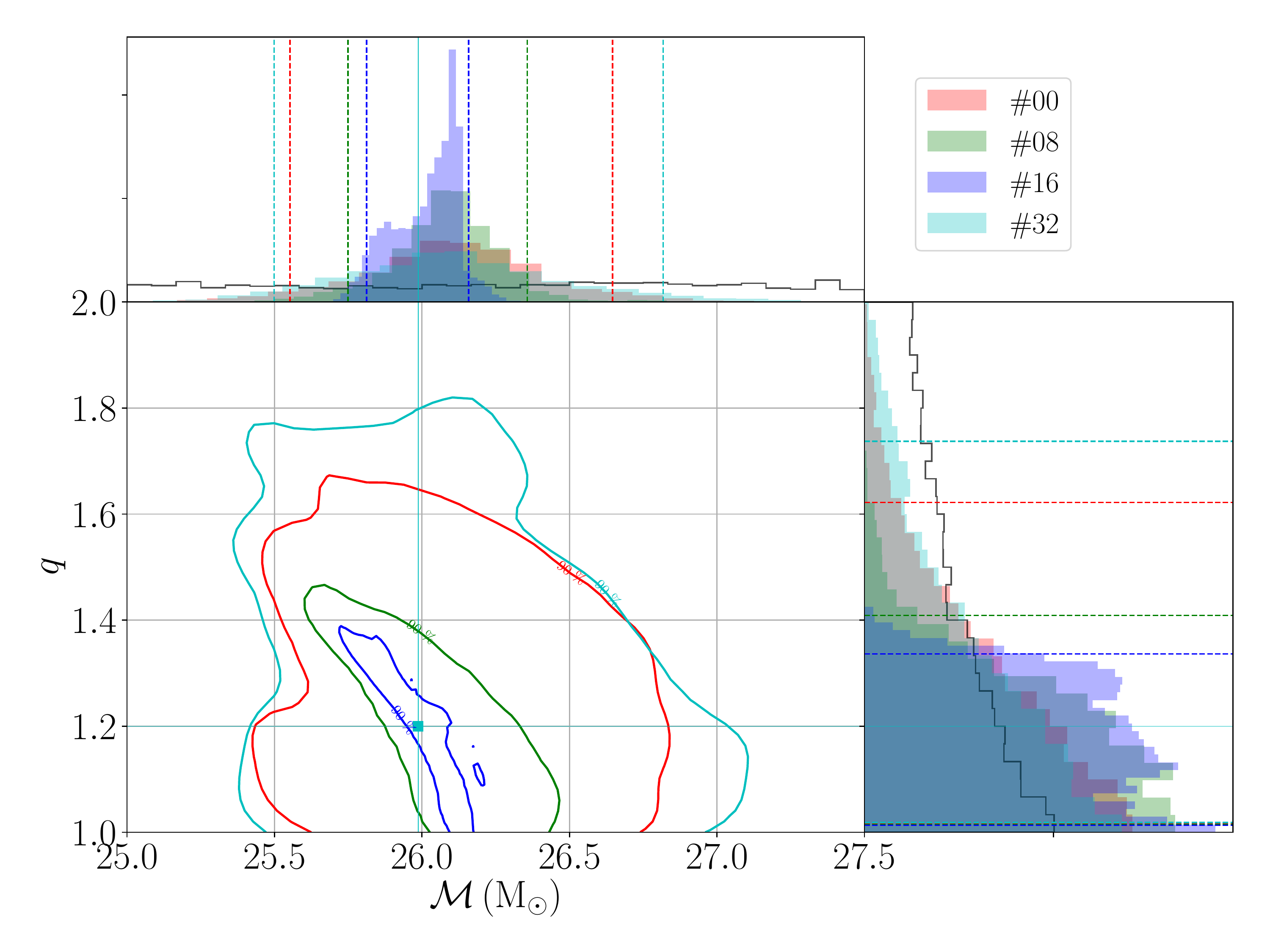}\\
 \includegraphics[width=0.97\columnwidth,clip=true,trim={0 0 0 0}]{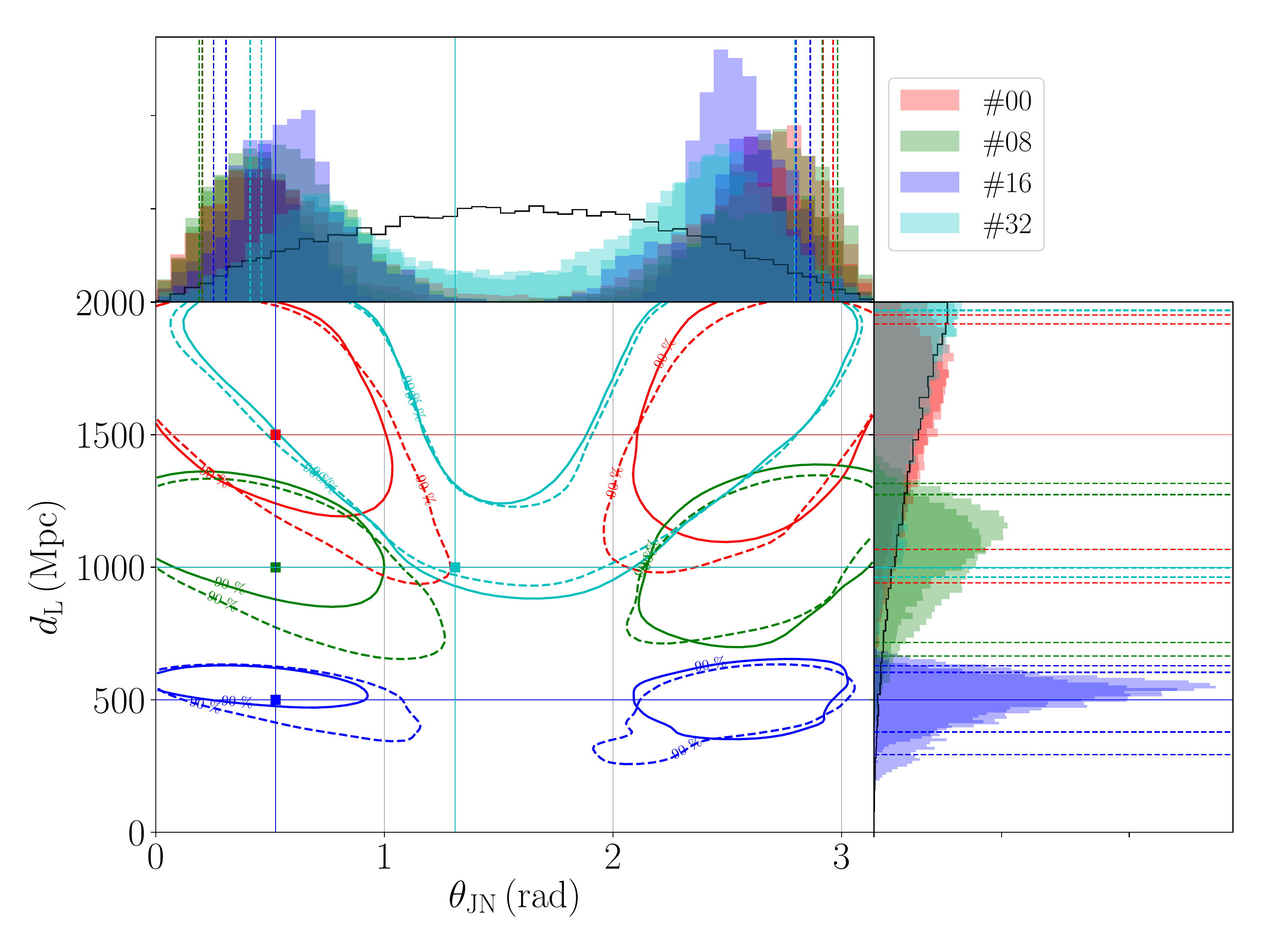}
 \caption{\label{fig:injections_masses_distincl}
 Parameter recovery for select synthetic injections with \NRSurHM{} templates.
 Here we show the mass (top) and source distance / inclination angle (bottom)
 recovery for four injections, labeled $\# 0, 8, 16, 32$ in
 Table.~\ref{tab:injections005}. Both top and bottom sub-figures have three
 panels: one that shows two dimensional $90\%$ credible regions for the
 joint measurement of both parameters (for that sub-figure), while the other two
 show one dimensional marginalized probability distributions measured for each
 of the same two source parameters (for that sub-figure).
 All $4$ injections have identical source mass and spin parameters.
 The first three have identical source inclination angles
 as well, but differ in the distance at which their source is located:
 $1500$Mpc, $1000$Mpc, and $500$Mpc respectively. The fourth injection
 ($\#32$) is similar to injection $\#8$ except that its orbital inclination
 is much closer to edge-on, i.e. $\tjn=75^\circ$ for $\#32$.
 In the bottom sub-figure, we additionally show results
 from analyses with \NRSurL{} templates as
 dashed $2$D contours. These can be directly contrasted with solid contours
 to read the effect of including $l>2$ modes in templates.
 In both $2$D and $1$D panels, solid colored lines mark the true injected
 parameter value, and dashed vertical lines show the limits of $90\%$ credible
 regions for the relevant parameter. In all $1$D panels, black curves
 show the sampling prior for that parameter.
 See text for further discussion.}
\end{figure}
\begin{figure*}
\includegraphics[width=0.99\textwidth,clip=true,trim={0 5mm 0 0}]{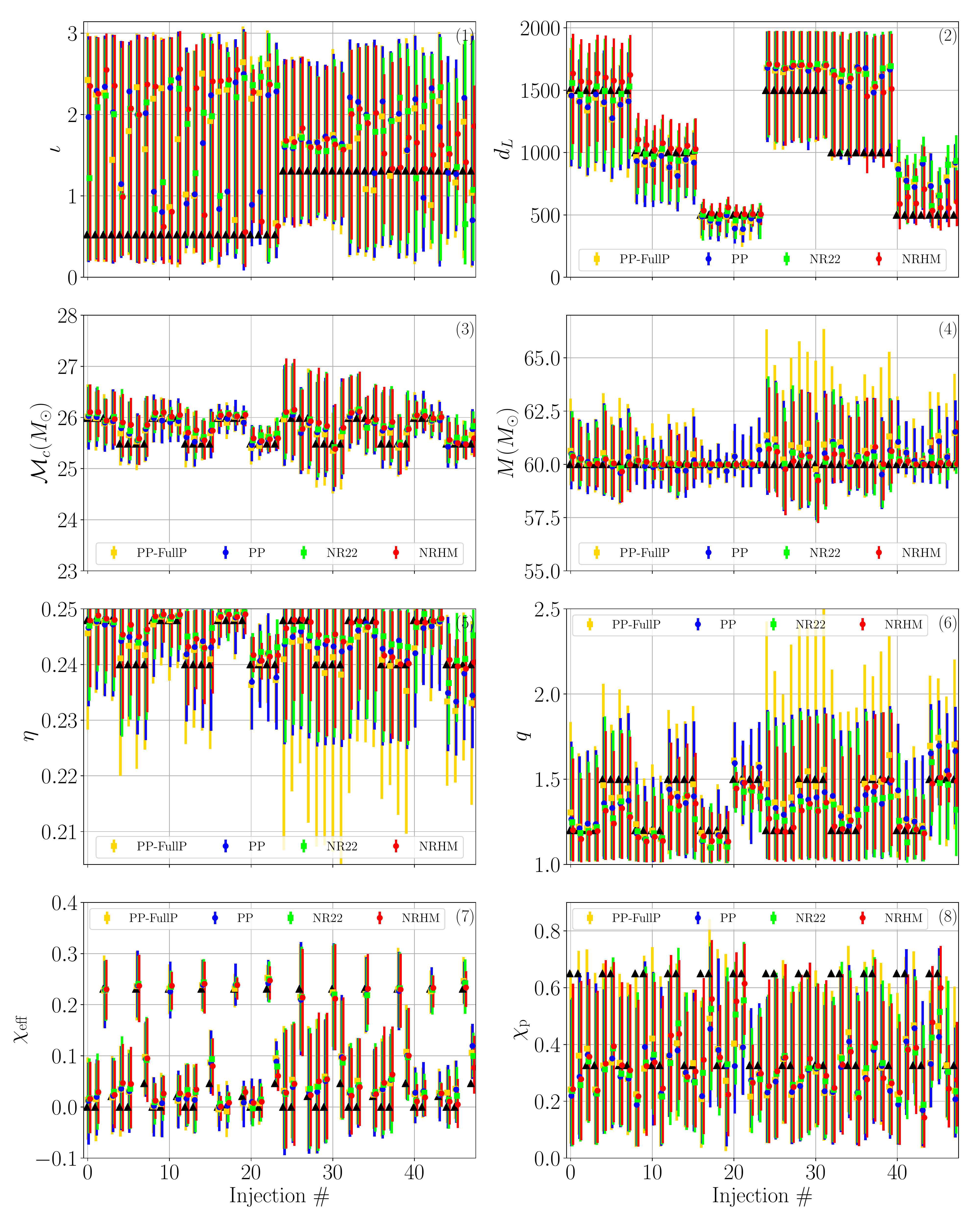}
 \caption{Estimated masses, spins, and other physical parameters for a set 
 of synthetic GW signals, analyzed using different approximants. In
 different colors, we show results for three template models: \Phenom{} (blue,
 labeled PP), \NRSurL{} (labeled NR22) and \NRSurHM{} (labeled NRHM), each
 restricted to the domain of validity of the \NRSur{} surrogate model. In
 addition we show results for \Phenom{} with the model allowed to explore
 its entire domain of support~\cite{Schmidt2010, Schmidt:2014iyl,
 Khan:2015jqa} (yellow, labeled PP-FullP).
 Parameters of injected signals are given in Table~\ref{tab:injections005}
 and described in text. True values of injection
 parameters are marked by black triangles. Vertical line segments
 show measured $90\%$ credible intervals, and colored circles show the
 corresponding median estimates.
 Panel numbers are indicated in top right corners. See text for discussion.
 \label{fig:m1_m2_all_limprior_injections005}}
\end{figure*}
%

Against these synthetic GW signals, we infer posterior probability
distributions for source parameters as described in
Sec.~\ref{s2:bayesian_methods}. For a pedagogical
overview, we start with examining a few select injections and study their
parameter recovery with \NRSur{}.
We choose $3$ injections corresponding to the
same binary, with $M=60M_\odot$, $q=1.2$, $\vec{\chi}_1 = -\vec{\chi}_2 =
\frac{0.65}{\sqrt{2}}(1,1,0)$, located at distances of $1500$Mpc, $1000$Mpc,
and $500$Mpc.
All of these $3$ are inclined to the line of sight at an angle $\tjn=30^\circ$.
These are labelled $\# 0,8,16$ in Table~\ref{tab:injections005}.
We also consider a fourth injection (labelled $\# 32$) that has the same
physical parameters as injection $\#8$, {\it but} is inclined at
$\tjn=75^\circ$. For all four, we show the recovery of their mass parameters
in the top sub-figure of Fig.~\ref{fig:injections_masses_distincl}, and their
orbital inclination and luminosity distance in the bottom sub-figure. For each
case, solid vertical lines in all panels showing $1$D histograms,
as well solid vertical and horizontal lines in $2$D panels, indicate
true injected parameter values. Dashed vertical lines indicate
measured bounds on them in the form of inferred $90\%$ credible intervals.
Let us focus first on the top sub-figure of this figure that focuses on 
chirp mass $\mchirp$ and mass ratio $q$. Looking first at $\mchirp$
recovery for the first $3$ cases, we notice a stark reduction in the width of
measured $90\%$ credible intervals with increasing SNR (or decreasing distance).
The measurement of $q$ also improves as the source moves closer from
$1500$ to $500$Mpc, albeit more slowly than for chirp mass.
The fourth injection ($\#32$, shown in cyan) is nearly edge-on with respect to
the line of sight to LIGO detectors.
Comparing it with the others, we immediately see how increasing the
source's inclination angle toward $\frac{\pi}{2}$
makes the measurement of BH masses significantly
worse. This is because $\tjn\rightarrow\frac{\pi}{2}$ decreases the
contribution of dominant $\ell=|m|=2$ modes, and therefore reduces the
overall SNR.
Further -
in the bottom sub-figure of Fig.~\ref{fig:injections_masses_distincl}, we show the
recovery of sources' inclination and luminosity distance from LIGO detectors.
All presentation attributes of this sub-figure are identical to those of the top,
with one addition. In panels showing $2$D credible regions, while
solid contours still correspond to \NRSurHM{}, we have added corresponding
{\it dashed} contours for \NRSurL{}.
We immediately see that the first three injections, which are nearly face-on,
have similar $90\%$ credible regions - each two-lobed around face-on and face-off
orientations. This is as we expect since both orientations are degenerate
and maximize the contribution of dominant $\ell=|m|=2$ waveform modes.
We find that
the presence of $l>2$ modes in recovery templates restricts the
distance-inclination posterior further, as seen by comparing solid and
dashed $2$D contours. Regions of the posterior that 
underestimate luminosity distance are ruled out more aggressively near
the lobe around $\tjn\rightarrow\tjn^\mathrm{true}$ than around
$\tjn\rightarrow\pi-\tjn^\mathrm{true}$ for this fiducial binary.
We find that
this asymmetry between face-on/face-off posterior lobes for distance and
inclination also depends on the intrinsic parameters of the source, which
modulate the relative signal power content between $l=3$ and $l=2$ modes.
Further: $\tjn=75^\circ$ being close to $\pi/2$, we expect a systematic
overestimation of distance for case $\#32$ as the true value is
located toward the lower ``U'' end of the $d_L - \tjn$ degeneracy contours.
We find, accordingly, that distance for the fourth injection is indeed 
grossly overestimated, in contrast with the other three cases. This is
consistent with past results for highly inclined binaries~\cite{Abbott:2016wiq}.

So far we have illustrated select cases of parameter estimation with
the \NRSur{} surrogate, highlighting differences between dominant-mode and
higher-mode templates. Next we will investigate the improvements in BBH
parameter recovery brought upon by both (a) the presence of higher-order
modes, and (b) NR-level accuracy of merger modeling, in \NRSurHM{} 
templates. We will do so analyzing all injections together.
We will quantify our results using marginalized $1$D $90\%$
credible intervals as measures of statistical error, with the
estimated median values furnishing any corresponding systematic errors.
Our results are shown in Fig.~\ref{fig:m1_m2_all_limprior_injections005} for
all injections. In each panel, the horizontal axis shows the injection
index, which ranges from $0-47$, and was introduced in
Table~\ref{tab:injections005}. Further, black triangles show injected
(true) parameter values.
Injections are arranged first according to their inclination
angle, then according to their source distance, then mass ratio, and
finally by the bigger BH's spin magnitude. This implies that the first $24$
injections shown have source $\tjn = 30^\circ$ and the next $24$ have $\tjn =
75^\circ$. Within each of these two blocks of $24$ injections, the first $8$
have sources at $d_L=1500$Mpc, next $8$ at $d_L=1000$Mpc and the last $8$
at $d_L=500$Mpc. Within each of these blocks of $8$ injections, the first
$4$ have $q=1.2$ and the next $4$ have $q=1.5$. And finally, within each
block of $4$ injections, the first $2$ have $|\vec{\chi}|_1=0.65$ and the
next $2$ have $|\vec{\chi}|_1=0.4$. This arrangement is manifest in the
locations of black triangles in all panels. The median value of the
measured marginalized posterior distribution for each parameter is shown
in solid circles and the associated $90\%$ credible intervals are shown as
vertical line segments. Colors distinguish between template models. Labels
NRHM and NR$22$ correspond to \NRSurHM{} and \NRSurL{} templates. Two sets of
results are shown with \Phenom{}, one where its sampling priors are
artificially restricted to the domain of \NRSur{} (labeled PP), and the
other where they span the entire domain of validity of \Phenom{} (labeled
PP-FullP).

In the top row of Fig.~\ref{fig:m1_m2_all_limprior_injections005}, panels
$(1)$ and $(2)$ show the recovery of source inclination and luminosity
distance with respect to LIGO detectors. We know that the effect of both
of these parameters on GW signals incident on Earth is degenerate, i.e., signals
with inclination $\tjn$ are degenerate with sources with $\tjn\rightarrow \pi-
\tjn$ at similar distances, as well as with sources with $\tjn\rightarrow
\pi/2$ at smaller distances or $\tjn\rightarrow 0$ at larger distances.
The general shape of this degeneracy is visually
appreciable from the $2$D cyan contours in the lower left corner of the
lower $2$D panel in Fig.~\ref{fig:injections_masses_distincl}.
Therefore, we find that median $\tjn$ estimates in panel $(1)$ of
Fig.~\ref{fig:m1_m2_all_limprior_injections005} are either close to the true
value of $\tjn$ or $\pi$ minus the true value. For most of nearly face-on
cases (with $\tjn=30^\circ$), neither \Phenom{} nor \NRSurL{} constrains binary's
initial orbital inclination very well, with $90\%$ credible intervals nearly
spanning the entire prior range $[0, \pi]$. For more inclined configurations
(i.e. with $\tjn = 75^\circ$), the effect of higher-order waveform modes
is enhanced, and we accordingly find that \NRSurHM{} constrains $\tjn$ better
than the other two models with only $\ell=|m|=2$ modes. This is especially
noticeable for sources closer than $1$Gpc (c.f. injections $\#32-47$).
In panel $(2)$ we show luminosity distance measurements. We notice
immediately that distance estimates can improve significantly for closer
sources with \NRSurHM{} templates.
For sources with $\tjn = 30^\circ$, we find that \NRSurHM{} measures source
distance more accurately and precisely than the other two models, with both
the median estimate being closer to the true value and $90\%$ credible
intervals being smaller.
For the same sources, we also find that while \NRSurL{} does not improve
the precision of luminosity distance measurement by much, it does improve
its accuracy, as the median estimates with \NRSurL{} are closer to true values
than with \Phenom{}. Contrasting the benefit of including higher-order
modes in \NRSurHM{} with improved merger modeling of \NRSur{},
we find the missing sub-dominant modes in \Phenom{} to be the leading
cause of loss of information in $d_L-\tjn$ measurements.
For closer to edge-on configurations (with
$\tjn=75^\circ$, injections $\#24-47$), we find that $d_L$ is {\it
systematically biased} toward larger values, which is expected given the
nature of $d_L-\tjn$ degeneracy, as we discussed above in the context of
Fig.~\ref{fig:injections_masses_distincl}. Even then, \NRSurHM{}
estimates are both more accurate and more precise for all injections,
especially note the loudest ones: $\#40-47$.
Finally, we point out that restricted priors for \NRSur{} do not reduce
the quality of either orbital inclination or luminosity distance
measurements with the model for chosen injections.

Next we look at the recovery of the binary's mass parameters. We show these
for four different mass combinations: chirp mass $\mchirp = M \eta^{3/5}$,
total mass $M$, dimensionless mass ratio $\eta = m_1 m_2 / M^2$, and
mass ratio $q$, in panels $(3)-(6)$ respectively.
For both chirp and total masses, we immediately notice that both are constrained
categorically worse for nearly edge-on cases than for nearly face-on ones.
For closer sources (out to $500$Mpc), we find that both \NRSurHM{} and
\NRSurL{} can measure both $\mchirp$ and $M$ more accurately and precisely
than \Phenom{} (especially total mass). The pattern holds for both
$\tjn=30^\circ$ and $75^\circ$ configurations.
A similar pattern is seen for the estimation of mass ratio $q$ (or
$\eta$\footnote{the one-to-one map: $\eta = q / (1+q)^2$ ensures that patterns
that hold for posteriors of $q$ will hold for $\eta$, and vice-versa.})
in panels $(5)-(6)$.
Overall, mass ratios are recovered better for nearly face-on systems.
For close binaries (out to $500$Mpc) we again find that mass ratios can be
better estimated by  both \NRSurHM{} and \NRSurL{} templates than with
\Phenom{} ones. This is especially evident for injections $\#16-23$ and
$\#40-47$.
These improvements clearly illustrate one benefit of the NR-based surrogate
that does not simply depend on the inclusion of extra information through
sub-dominant GW modes, but on its intrinsic ability to reproduce GR signals
more faithfully. We therefore surmise that, as expected, {\it the benefit of
following up GW signals with NR surrogate templates is genuinely twofold}.
We do note, however, that for highly inclined configurations, the 
artificially restricted prior of \NRSur{} shows up as the leading factor
affecting the quality of $M,\eta, q$ measurements (but not of $\mchirp$).
This is seen by comparing the two set of results with \Phenom{} (PP and
PP-FullP) for injections $\#24-39$ in panels $(3)-(6)$. This clearly
motivates the development of NR-based surrogate models for more unequal
mass ratios.

\begin{figure*}
\centering
 \includegraphics[width=0.98\columnwidth,clip=true,trim={2mm 6mm 4mm 0}]{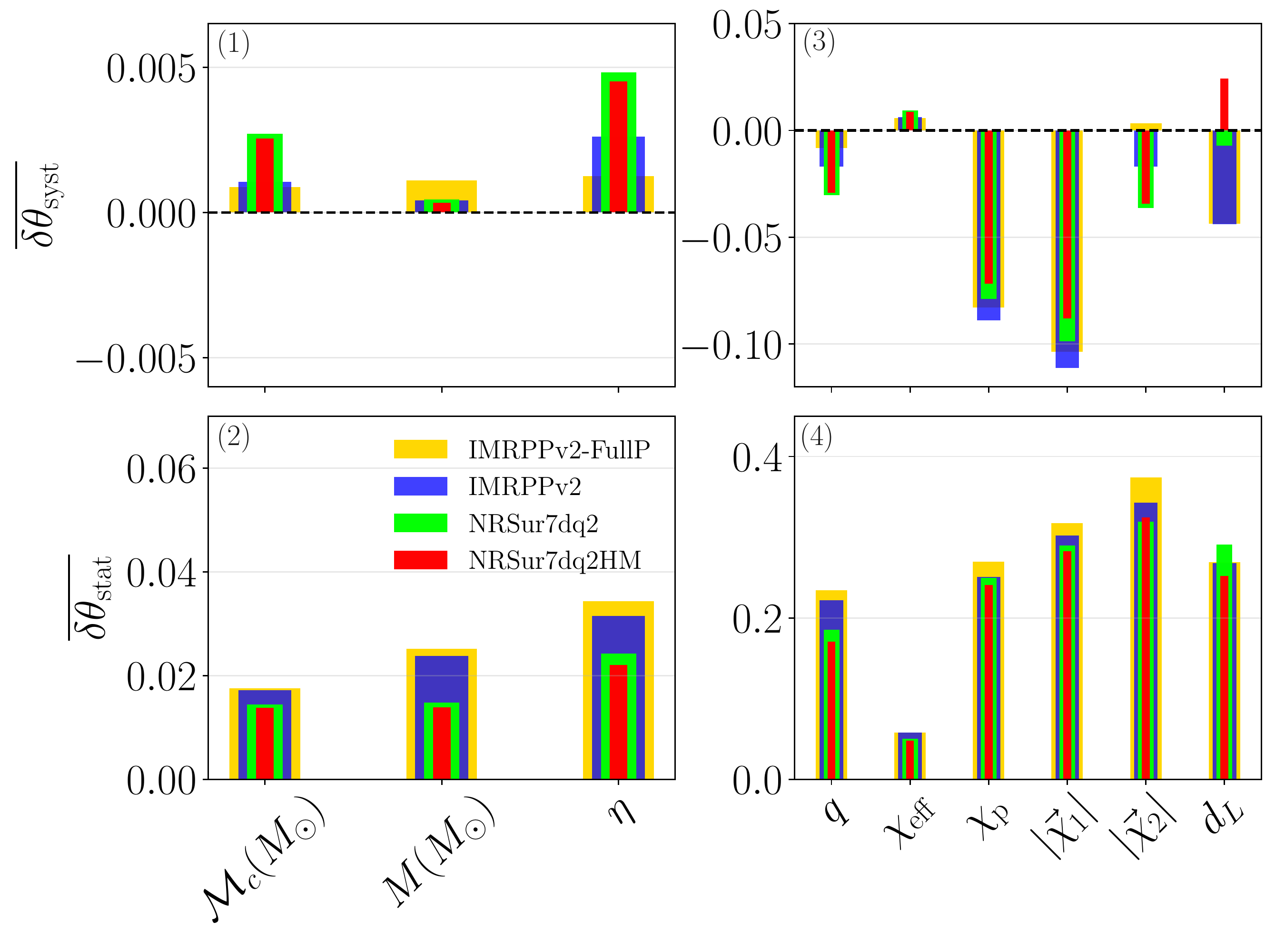}
 \includegraphics[width=1.06\columnwidth,clip=true,trim={2mm 5mm 0mm 0}]{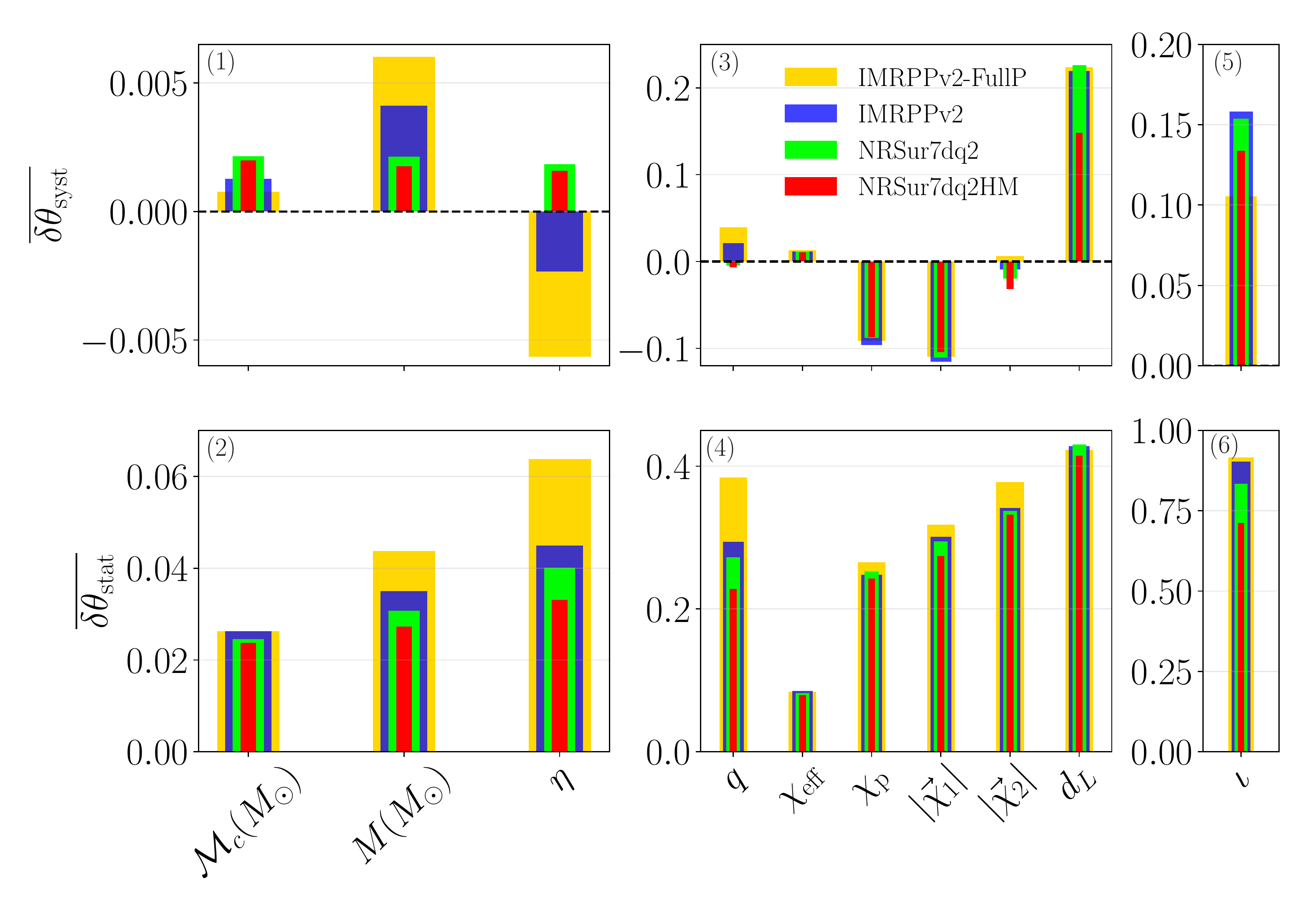}
 \caption{{\it Left}: Shown are the mean systematic biases 
 \tsyst{} and statistical uncertainties \tstat{} for various binary parameters
 averaged over all injections with $\tjn=30^\circ$. Four distinct template
 configurations are considered: \NRSurL{} and \NRSurHM{} models, and
 \Phenom{} with and without being artificially restricted to the domain
 of validity of \NRSur{}.
 The full prior for \Phenom{} extends over $1\leq q\leq 8$ and spin
 magnitudes $|\vec{\chi}_{1,2}|\leq 0.89$.
 Since the signal model is \NRSurHM{}, when the recovery model is also
 \NRSurHM{} the mean systematic biases and statistical uncertainties reflect
 the shape of the posterior itself rather than modeling error.
 {\it Right}: Same as left panel, except the averaging of \tsyst{} and \tstat{}
 is performed over all injections with $\tjn=75^\circ$.
 \label{fig:all_param_mean_errors_injections005}}
\end{figure*}
%

Finally, we consider BH spins. We show the recovery of two different 
combinations of both BH's spins in panels $(7)-(8)$ of
Fig.~\ref{fig:m1_m2_all_limprior_injections005}: the effective spin $\chieff$,
and the in-plane precessing spin $\chi_p$.
From panel $(7)$ we note that effective spin is consistently well
estimated by all template models. For the closest sources at $500$Mpc,
we find that $\chieff$ is estimated more {\it precisely} by
both \NRSurHM{} and \NRSurL{}, i.e. with narrower $90\%$ credibility
intervals, than other approximants; see, for instance, injections $\#8-23$
and $\#40-47$.
This (marginal) improvement comes from the
more faithful modeling of spin effects in the dominant GW mode of
\NRSur{}, as it does not depend on the inclusion of higher-order modes.
From panel $(9)$, we note that $\chi_p$ is overall
poorly constrained for heavy BBHs such as the cases considered here. The
measured $90\%$ credible regions almost span $90\%$ of the entire prior
range, implying that very little information about $\chi_p$ is available.
This is to be expected because the timescale of orbital precession is
considerably larger than the orbital timescale. Short signals from
heavy BBHs barely span a couple of precession cycles, making measurements
of precession-related parameters challenging. We still note, however,
that for the closest injected sources at moderate inclination
($\tjn = 30^\circ$, i.e. injections $\#16-23$), both \NRSurHM{} and
\NRSurL{} recover $\chi_p$ somewhat more accurately than \Phenom{}.
For these spin measurements, we again find no substantial influence of
artificially restricted priors of \NRSur{} models.

A more succinct way of summarizing information from all injections is to
compute an averaged measure of systematic biases and statistical uncertainties
associated with the recovery of various physical parameters $\theta$ by
different template models. For each parameter we therefore first
compute the relative
systematic bias $\delta\theta^i_\mathrm{syst}$ and relative statistical
uncertainty $\delta\theta^i_\mathrm{syst}$ for each injection (indexed
$\#i$) as
\begin{eqnarray}\label{eq:bias_defs}
 \delta\theta^i_\mathrm{syst} &:=& \left|\theta^i_\mathrm{median} - \theta^i_\mathrm{true} \right| / \theta^i_\mathrm{true}, \\ \nonumber
 \delta\theta^i_\mathrm{stat} &:=& \left|\Delta\theta^i_\mathrm{90\%}  \right| / \theta^i_\mathrm{true},
\end{eqnarray}
where $\Delta\theta^i_\mathrm{90\%}$ is the size of the measured
$90\%$ credible region.
For parameters whose possible values include $0$, such as BH spins and their 
combinations, we do not divide by $\theta^i_\mathrm{true}$ in both parts of
Eq.~\ref{eq:bias_defs}.
We then take the algebraic mean of both $\delta\theta^i_\mathrm{syst}$
and $\delta\theta^i_\mathrm{stat}$ over all injections to obtain our
combined measures of parameter estimation accuracy and precision: \tsyst{} and
\tstat{}. We find that both of these measures are significantly affected by 
the inclination angles of injections being averaged over. We therefore average
over two sets of injections separately, one where the injected $\tjn=30^\circ$
and another where $\tjn=75^\circ$.
We show summary error measures for both sets of injections in the two
sub-figures of Fig.~\ref{fig:all_param_mean_errors_injections005}. The left
sub-figure corresponds to injections with $\tjn=30^\circ$ and the right one
corresponds to those with $\tjn=75^\circ$. We show results
for \NRSurL{}, \NRSurHM{}, and two configurations of \Phenom{} templates
(using unrestricted and \NRSur{}'s restricted sampling priors, labeled IMRPPv2
and IMRPPv2-FullP respectively). As in
Fig.~\ref{fig:m1_m2_all_limprior_injections005}, colors indicate different
template models.

Let us focus on the left sub-figure of
Fig.~\ref{fig:all_param_mean_errors_injections005}.
Considering the recovery of binary mass combinations first, we immediately
note that for $\mchirp, M, \eta, q$: \tsyst{} $\ll$ \tstat{}, and
therefore statistical errors dominate their measurement.
By comparing results from different template models, we can see that {\it both}
\NRSurHM{} and \NRSurL{} recover all four mass combinations
substantially more {\it precisely} than \Phenom{} templates. Most noticeable
is the improvement in measuring total mass. This improvement is unlikely
to be due to artificially restrictive sampling priors of \NRSur{},
as the effect of the same priors is minimal on \Phenom{} analyses
(as can be seen by comparing blue and yellow bars in panel $(2)$).
We therefore conclude that {\it the improved modeling of dominant GW mode
by} \NRSur{} (in the nonlinear merger regime) {\it is responsible
for this improvement in our capability to measure BBH masses}.
Next, we turn our attention to BBH spin combinations $\chieff, \chip,
|\vec{\chi}_1|,$ and $|\vec{\chi}_2|$. Results for these are shown in 
panels $(3)-(4)$. We again note that the ratio \tsyst{}/\tstat{} is below
$10\%$ for all four, implying that statistical errors dominate.
From panel $(3)$, we read that while for $\chip$ and
$|\vec{\chi}_1|$ the surrogates record smaller systematic measurement biases,
for $\chieff$ and the smaller BH's spin its the opposite. However,
this improvement is moot unless we improve on the dominant statistical errors.
We therefore turn to panel $(4)$. We find that while both \NRSur{}
models
slightly improve the precision of measurement for individual BH spins and
$\chip$, this improvement is contaminated by the restricted
sampling priors of \NRSur{}. This can be seen by comparing yellow and blue
bars for the three spin combinations in panel $(4)$. Having said this,
we remind the reader that for these spin combinations, little
information is actually recovered from data as measurements tend to follow
sampling priors for such heavy BBHs. Lastly, we find a
small improvement in the measurement precision for effective spins
with \NRSurL{}({\tt HM}), which is too marginal for us to draw generic
conclusions. Looking back at panel $(7)$ of
Fig.~\ref{fig:m1_m2_all_limprior_injections005} we remind the reader that
\NRSurHM{} improves the measurement of $\chieff$ only for the
closest sources (out to $500$Mpc), and this improvement gets washed out
when we average over all other injections at $1000-1500$Mpc.
Finally, we assess the measurement quality for BBH luminosity distances.
Looking at panel $(4)$, we find that both surrogate and \Phenom{} templates
measure $d_L$ with comparable {\it precision}. From the panel right above, 
however, we find that \NRSurHM{} systematically over-estimates $d_L$ by
only about $2.5\%$ while \Phenom{} under-estimates it by twice that
amount.
We also find that while the former systematically overestimates
$d_L$, the latter underestimates it. This can be understood heuristically
by considering the shape of $d_L-\tjn$ degeneracy contours. Look back
at the illustrations in Fig.~\ref{fig:injections_masses_distincl}, and
focus on the lower sub-figure. The $2$D contours show that \NRSurHM{} improves
upon the measurement with \NRSurL{} by ruling out the lower portions
of the $d_L-\tjn$ degeneracy regions. These portions correspond to 
highly-inclined close-by configurations. Since \NRSurHM{} eliminates
them from the posterior, the resulting posterior is bound to move
toward larger values of $d_L$. This can be seen by comparing any of
the red/green/blue solid contours with dashed ones in the same $2$D panel.
This explains why \tsyst{} for $d_L$ is strictly positive with \NRSurHM{},
but is negative with the dominant-mode-only models.
However, since the ratio \tsyst{}/\tstat{} for $d_L$ is $10\%$
with \NRSurHM{}, $4\%$ for \NRSurL{}, and $16\%$ for \Phenom{},
statistical errors still dominate the measurement of $d_L$ {\it on average}.
Reconciling this observation with panel $(2)$ of
Fig.~\ref{fig:m1_m2_all_limprior_injections005} (injections $\#16-23$),
we conclude that \NRSurHM{} can improve the {\it accuracy} of $d_L$ measurement
provided the GW source is close enough ($\lesssim 500$Mpc).
Note that we do not discuss the measurement quality for $\tjn$ itself
as we find to be nearly identical between all template models.

Moving forward, we focus on nearly edge-on injections in the right sub-figure of 
Fig.~\ref{fig:all_param_mean_errors_injections005}. Compared to the
nearly face-on cases, we immediately note that the measurement of mass
parameters including $M, \eta,$ and $q$ by \NRSur{} are uncertain
enough to be dominated by the restriction on the models' domain of validity.
In other words, the full posterior distributions for these parameters have
substantial support outside the domain of \NRSur{}
and therefore these NR
surrogate models will likely produce a biased estimate for them. Having said
this, we also note that the chirp mass is measured fairly consistently with all
models. It is measured with high precision, and its recovered posteriors
are narrow enough to lie completely within the domain of \NRSur{} models,
unaffected by sampling prior restrictions.
Looking at spin parameters $\chieff, \chip, |\vec{\chi}_1|,$ and
$|\vec{\chi}_|$ next, we find qualitatively very similar features as we did 
for nearly face-on injections in the left sub-figure of
Fig.~\ref{fig:all_param_mean_errors_injections005}. The measurement of
$\chieff$ is consistent between all four template choices, while that of
the other three spin combinations is mildly influenced by the the sampling
prior restrictions.
Lastly, we focus on the measurement of source distance and inclination.
From panels $(3)-(4)$ we find that while all four template models
measure luminosity distances to comparable precision, the inclusion of
sub-dominant waveform modes in \NRSurHM{} does improve its {\it accuracy}
substantially. Since \tsyst{}/\tstat{} is close to $50\%$ for $d_L$ here,
this improvement in \tsyst{} by \NRSurHM{} templates is substantial\footnote{%
Also note that for edge-on injections (right sub-figure of
Fig.~\ref{fig:all_param_mean_errors_injections005}), luminosity distance $d_L$
is systematically {\it over}-estimated by {\it all} models, as has been
found before~\cite{Abbott:2016wiq} (see Fig.~$4$)}. Turning to panels
$(5)-(6)$ we find a similar story. The inclusion of higher-order modes
in \NRSurHM{} again lead to a substantial reduction in both the
systematic and statistical errors associated with measuring
orbital inclination. Looking back at panel $(1)$ of
Fig.~\ref{fig:m1_m2_all_limprior_injections005} we confirm that this is
especially true for closer sources (out to $500$Mpc).

From these results, we conclude that parameter recovery with \NRSurHM{}
templates can be an improvement over conventional precessing template
model that have been used so far to analyze LIGO-Virgo BBH observations~\cite{
LIGOVirgo2016a,Abbott:2016nmj,TheLIGOScientific:2016pea,
Abbott:2017vtc,Abbott:2017oio,LIGOVirgo2018:GWTC_1}. These surrogate
templates can help estimate source masses better for moderately inclined
BBH configurations, and for comparable-mass close-by sources they
help resolve the luminosity distance-orbital inclination degeneracy and
improve the 
measurement of both. Our results emphasize the impact of two factors that
set NR surrogates apart from other models: (a) NR-level accurate
modeling of the dominant GW modes, and (b) inclusion of higher-order
harmonics. Measurement improvements that we find here due to (b)
are consistent with past work on higher harmonics~\cite{%
Graff:2015bba,London:2017bcn,PekowskyEtAl:2012,Healy:2013jza}, while
those due to $(a)$ are a novel result.
Based on these findings, we encourage the GW community to utilize NR
surrogates for detailed follow-ups of heavy BBH coalescences.
We also motivate the NR community to continue further development of
surrogate models, as extending their domain to higher mass ratios
can broaden the scope of their applicability.


\begin{figure}
\centering
 \includegraphics[width=\columnwidth,clip=true,trim={6mm 8mm 5mm 0mm}]{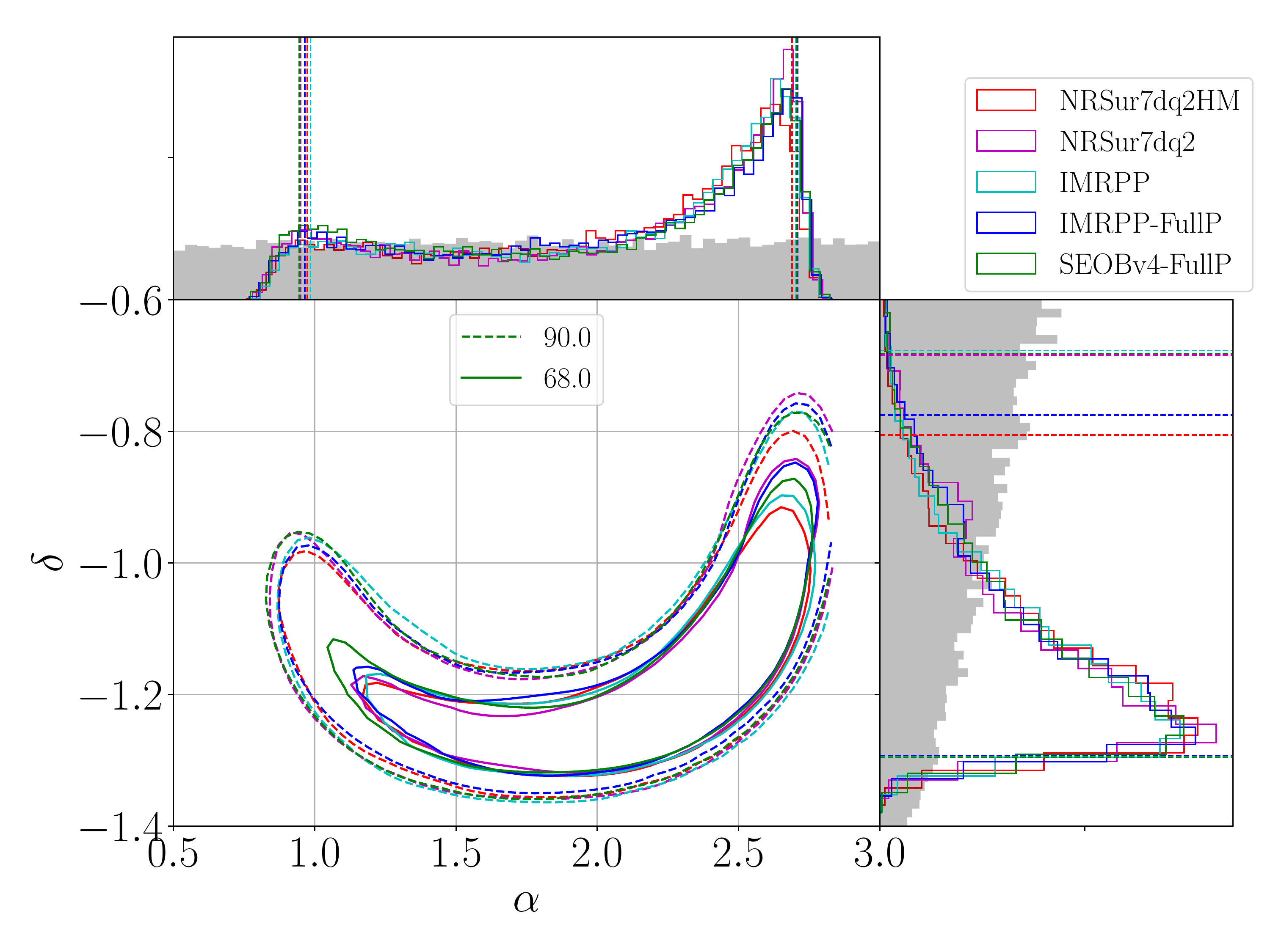}
 \caption{Estimated sky location for GW150914, using different approximants: 
 \NRSurHM{}, \NRSurL{}, \Phenom{} (labeled IMRPP), and \SEOB{} (labeled SEOBv4).
 The X-FullP results correspond to an analysis with model X that allows for
 unrestricted mass ratios $1\leq q \leq 8$, and spin magnitudes up to
 $a_{1,2}\lesssim 0.89$ for \Phenom{} and $a_{1,2}\lesssim 0.98$ for
 \SEOB{}. For all others, we {\it a priori} restrict
 sampling to $1\leq q \leq 2$ and $0\leq a_{1,2}\leq 0.8$, i.e. to the range where NR
 surrogate models are valid.
 In all panels showing $1$-D posterior distributions, the shaded region shows our 
 prior belief. Vertical dashed lines in $1$-D posteriors mark $90\%$ credible regions.
 The 2-D posteriors show both the $90\%$ (dashed line) and $68\%$ (solid line)
 credible regions.
 \label{fig:ra_dec_68_90pc_NR_NRHM_PP_PPFullP_gw150914}}
\end{figure}
\begin{figure}
\centering
 \includegraphics[width=\columnwidth,clip=true,trim={5mm 5mm 5mm 0mm}]{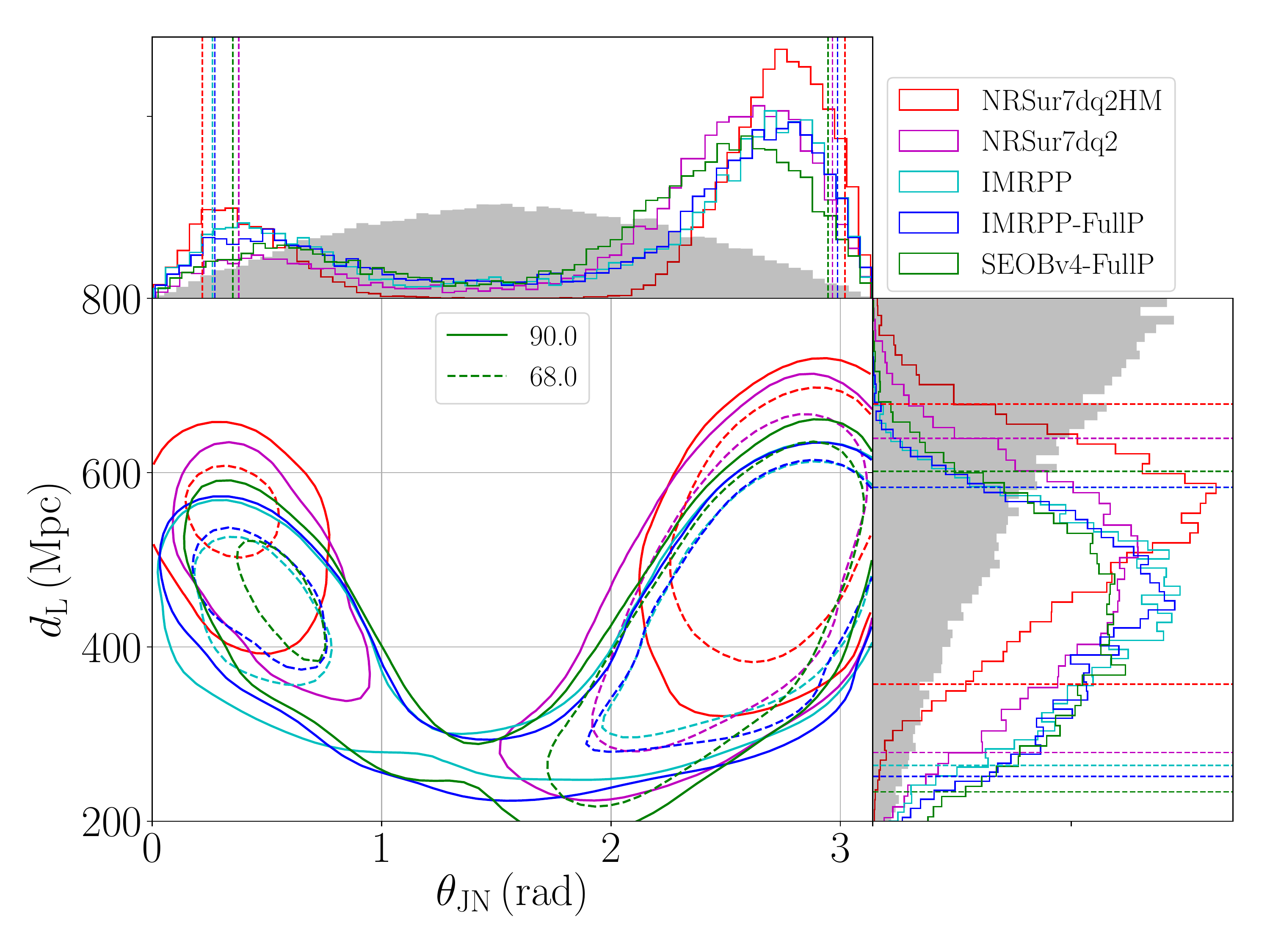}
 \caption{Estimated source orbital orientation / luminosity distance
 for GW150914, using different approximants: 
 \NRSurHM{}, \NRSurL{}, \Phenom{} (labeled IMRPP), and \SEOB{} (labeled SEOBv4).
 All figure attributes are similar to
 Fig.~\ref{fig:ra_dec_68_90pc_NR_NRHM_PP_PPFullP_gw150914}.
 \label{fig:dist_incl_68_90pc_NR_NRHM_PP_PPFullP_gw150914}}
\end{figure}
\begin{figure}
\centering
 \includegraphics[width=\columnwidth,clip=true,trim={5mm 5mm 0 0}]{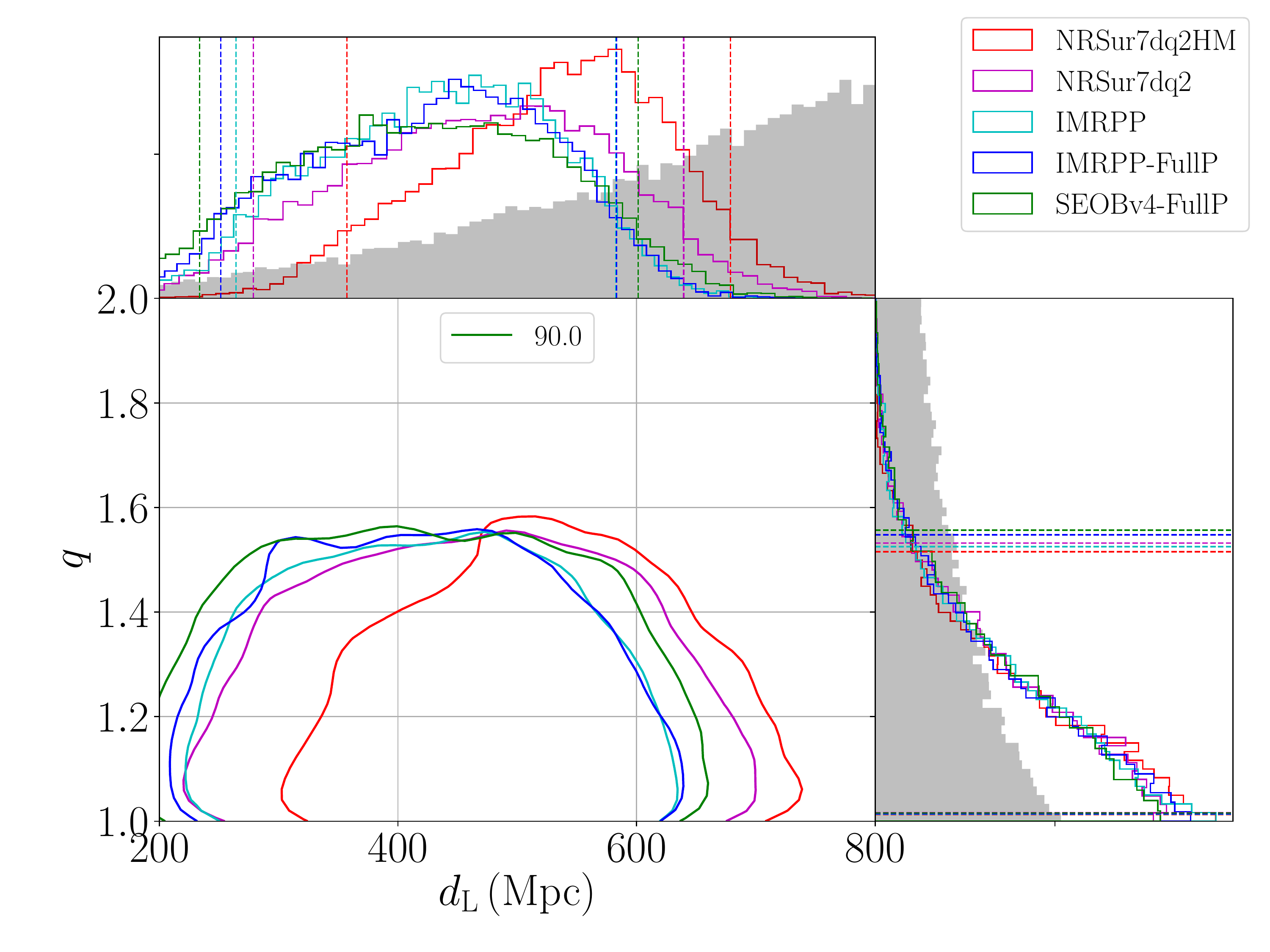}
 \caption{Luminosity distance and mass ratio measurement for GW150914.
 All figure attributes are similar to
 Fig.~\ref{fig:ra_dec_68_90pc_NR_NRHM_PP_PPFullP_gw150914}.
 We find that the samples at large luminosity distances actually 
 correspond to {\it smaller} mass ratios, and therefore the shifting
 of distance measurement to larger values when using \NRSurHM{} is
 not a symptom of the model's restricted sampling priors.
 \label{fig:dist_q_90pc_NR_NRHM_PP_PPFullP_gw150914}}
\end{figure}
\begin{figure*}
\centering
 \includegraphics[width=0.495\textwidth,clip=true,trim={8mm 5mm 0 0}]{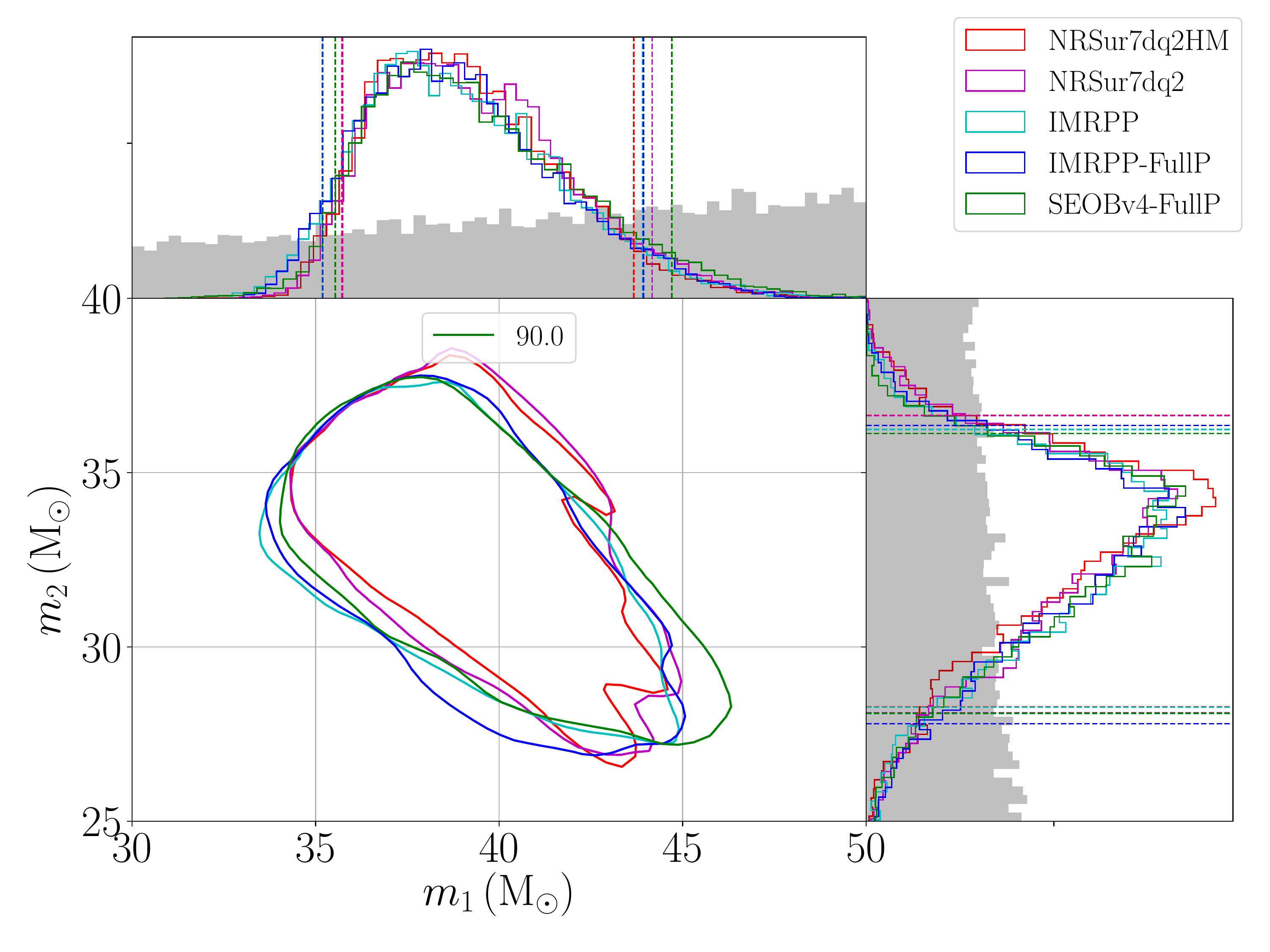}
 \includegraphics[width=0.495\textwidth,clip=true,trim={8mm 5mm 0 0}]{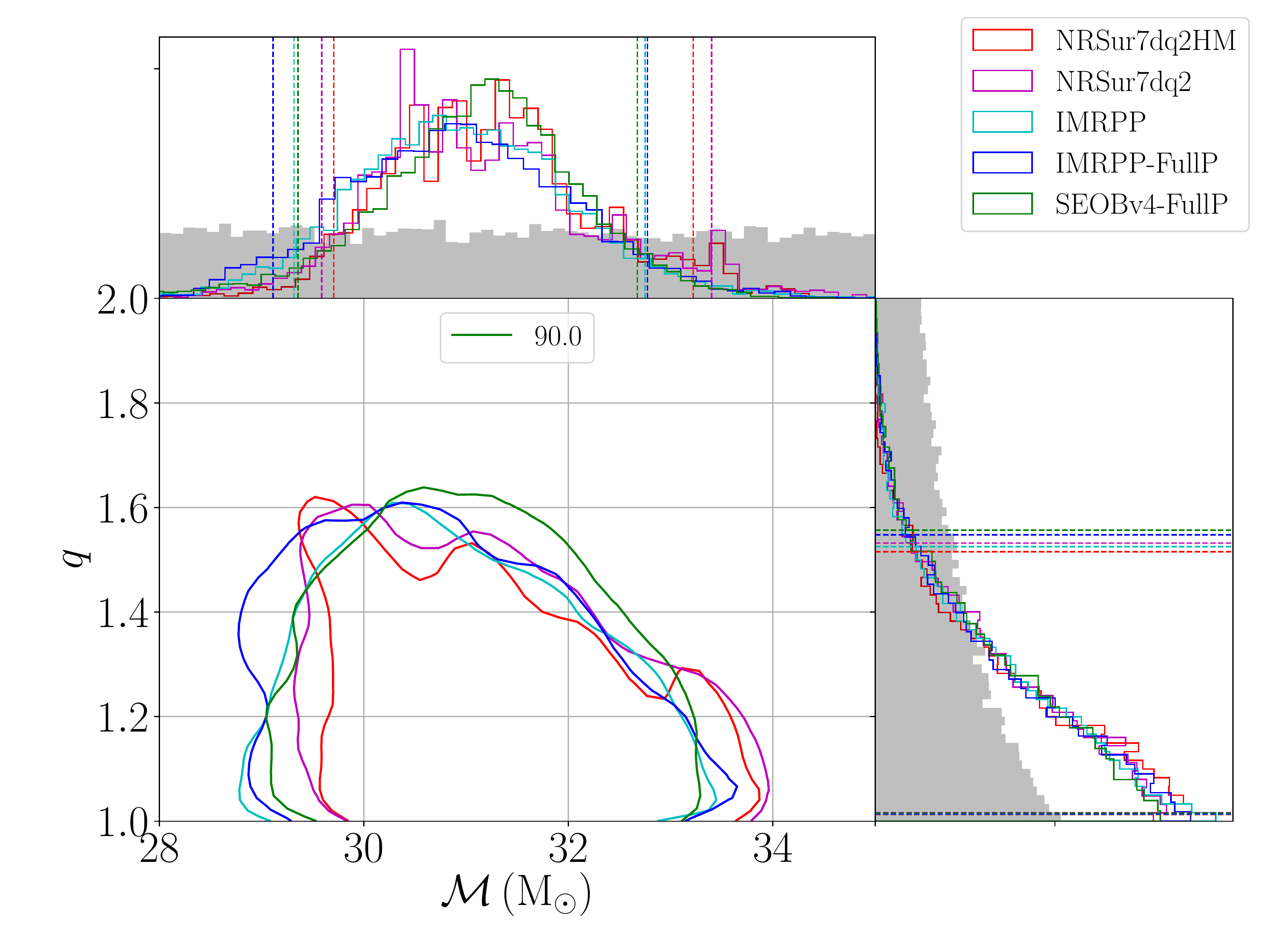}
 \caption{Estimated masses for GW150914, using different approximants: \NRSurHM{},
 \NRSurL{}, \Phenom{} (labeled IMRPP), and \SEOB{} (labeled SEOBv4).
 The X-FullP results correspond to an analysis with model X that allows for
 unrestricted mass ratios $1\leq q\leq 8$, and spin magnitudes up to
 $a_{1,2}\lesssim 0.89$ for \Phenom{} and $a_{1,2}\lesssim 0.98$ for
 \SEOB{}. For all others, we {\it a priori} restrict
 sampling to $1\leq q \leq 2$ and $0\leq a_{1,2}\leq 0.8$, i.e. to the range where NR
 surrogate models are valid.
 In all panels showing $1$-D posterior distributions, the shaded region shows our
 prior belief. Vertical dashed lines in $1$-D posteriors mark $90\%$ credible regions.
 The 2-D posteriors show the $90\%$ credible regions as a solid line.
 \label{fig:m1_m2_90pc_NR_NRHM_PP_PPFullP_gw150914}}
\end{figure*}
\begin{figure}
\centering
 \includegraphics[width=\columnwidth,clip=true,trim={5mm 5mm 0 0}]{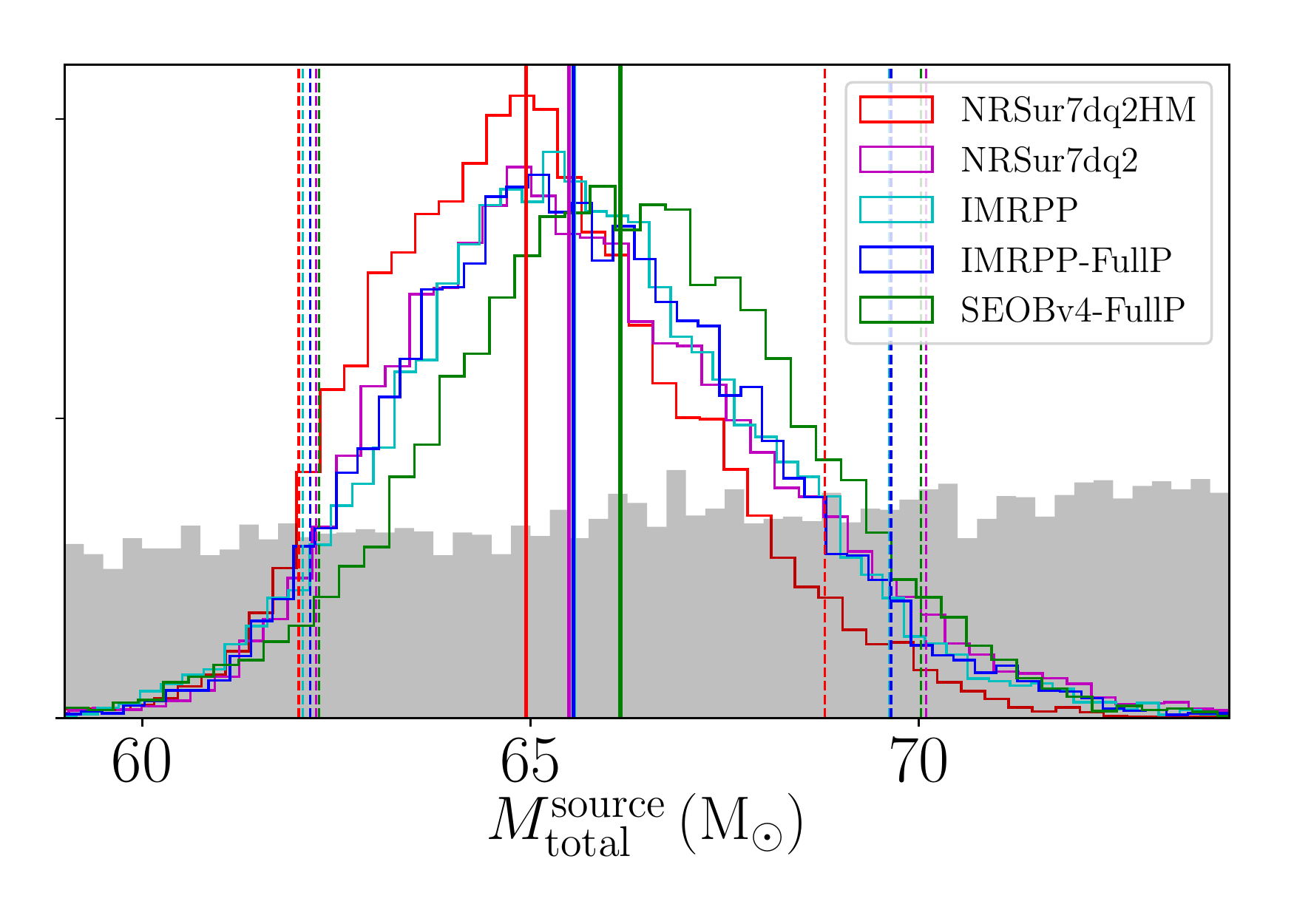}
 \caption{Estimated total mass for GW150914, as measured in the
 source frame. Four different approximants are shown: \NRSurHM{},
 \NRSurL{}, \Phenom{} (labeled IMRPP), and \SEOB{} (labeled SEOBv4).
 Figure attributes are identical to Fig.~\ref{fig:m1_m2_90pc_NR_NRHM_PP_PPFullP_gw150914}.
 The shaded region shows our prior belief. Vertical dashed lines mark $90\%$
 credible regions, and vertical solid lines show the distribution median.
 \label{fig:mtot_source_90pc_NR_NRHM_PP_PPFullP_gw150914}}
\end{figure}
\begin{figure*}
\centering
 \includegraphics[width=0.495\textwidth,clip=true,trim={8mm 5mm 0 0}]{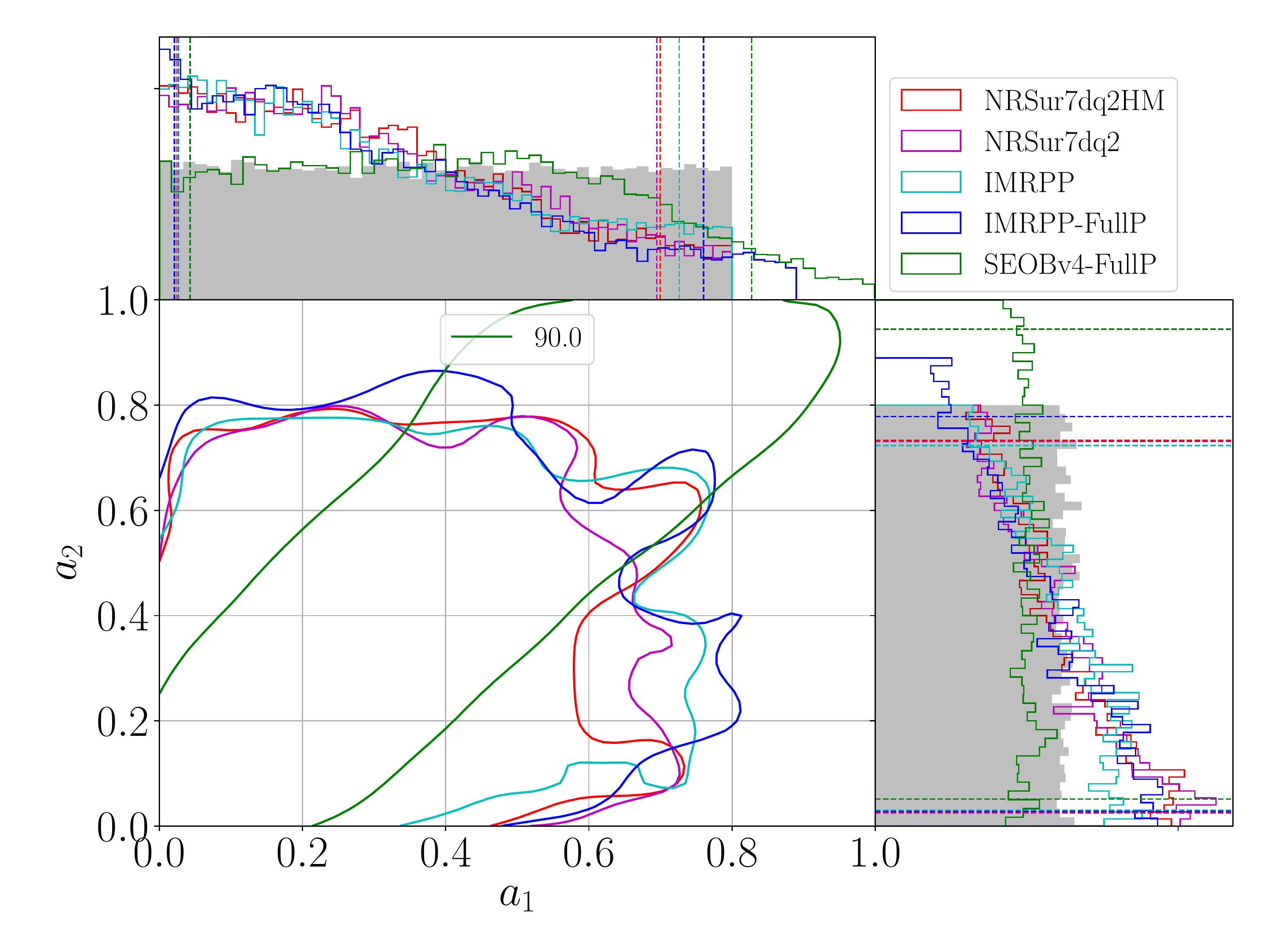}
 \includegraphics[width=0.495\textwidth,clip=true,trim={8mm 5mm 0 0}]{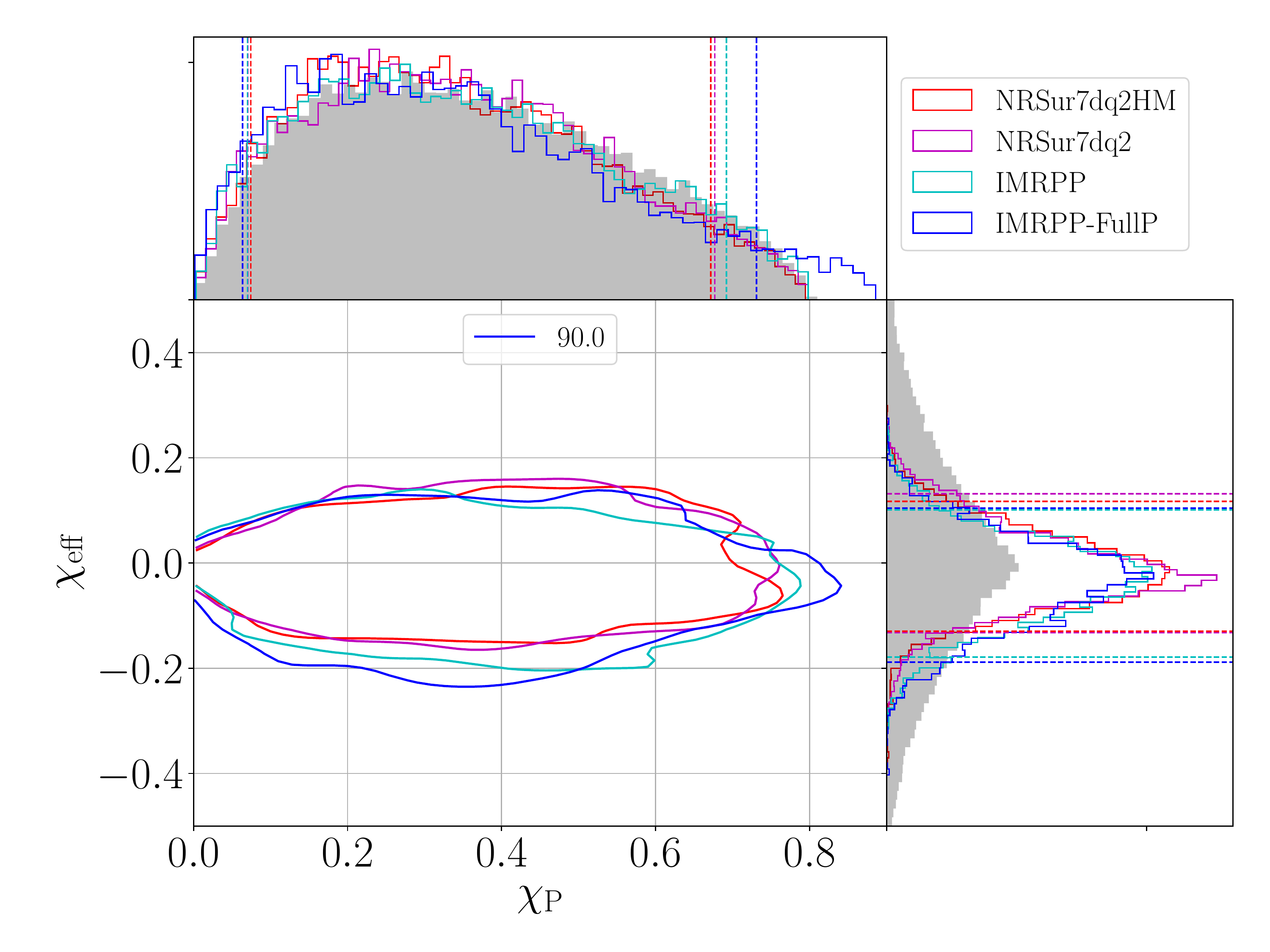}
 \caption{Estimated spins for GW150914, using different approximants and
 different prior probability distributions. Shown are spin magnitudes for
 both BHs and the tilt angles between BH spins and the orbital angular
 momentum at $f_\mathrm{ref}$. Figure attributes are identical to 
 Figs.~\ref{fig:ra_dec_68_90pc_NR_NRHM_PP_PPFullP_gw150914}
 and~\ref{fig:m1_m2_90pc_NR_NRHM_PP_PPFullP_gw150914}.
 \label{fig:a1_a2_tilt1_tilt2_90pc_NR_NRHM_PP_PPFullP_gw150914}}
\end{figure*}
%

We remind the reader that the choice
of injected parameters here is made to enhance the effect of precession, and so
could be considered a sample of ``moderately'' precessing sources.
It is not, however, drawn from an astrophysically motivated distribution, and
is therefore {\it not} representative of an astrophysical BBH
population (in any case, a sample size of $48$ over an $8$D parameter space
is unlikely to be {\it statistically} representative of any chosen
distribution). Therefore, knowing how much benefit we will reap with
NR surrogates for a (future) LIGO-Virgo BBH population
would require the additional knowledge of how the source parameters of
LIGO-Virgo BBH sources are distributed in nature;
a study of this is beyond the scope of this article.
Finally, note that the choice of using
zero noise instead of a particular noise realization ensures that
our results hold {\it on average}, where the averaging is meant in the sense of
an ensemble average over an infinite set of noise realizations embedding the
same signal. When real instrument noise is present, these results will 
get shifted depending on the exact nature of the noise realization.



\begin{figure}
\centering
 \includegraphics[width=\columnwidth,clip=true,trim={8mm 8mm 0 0}]{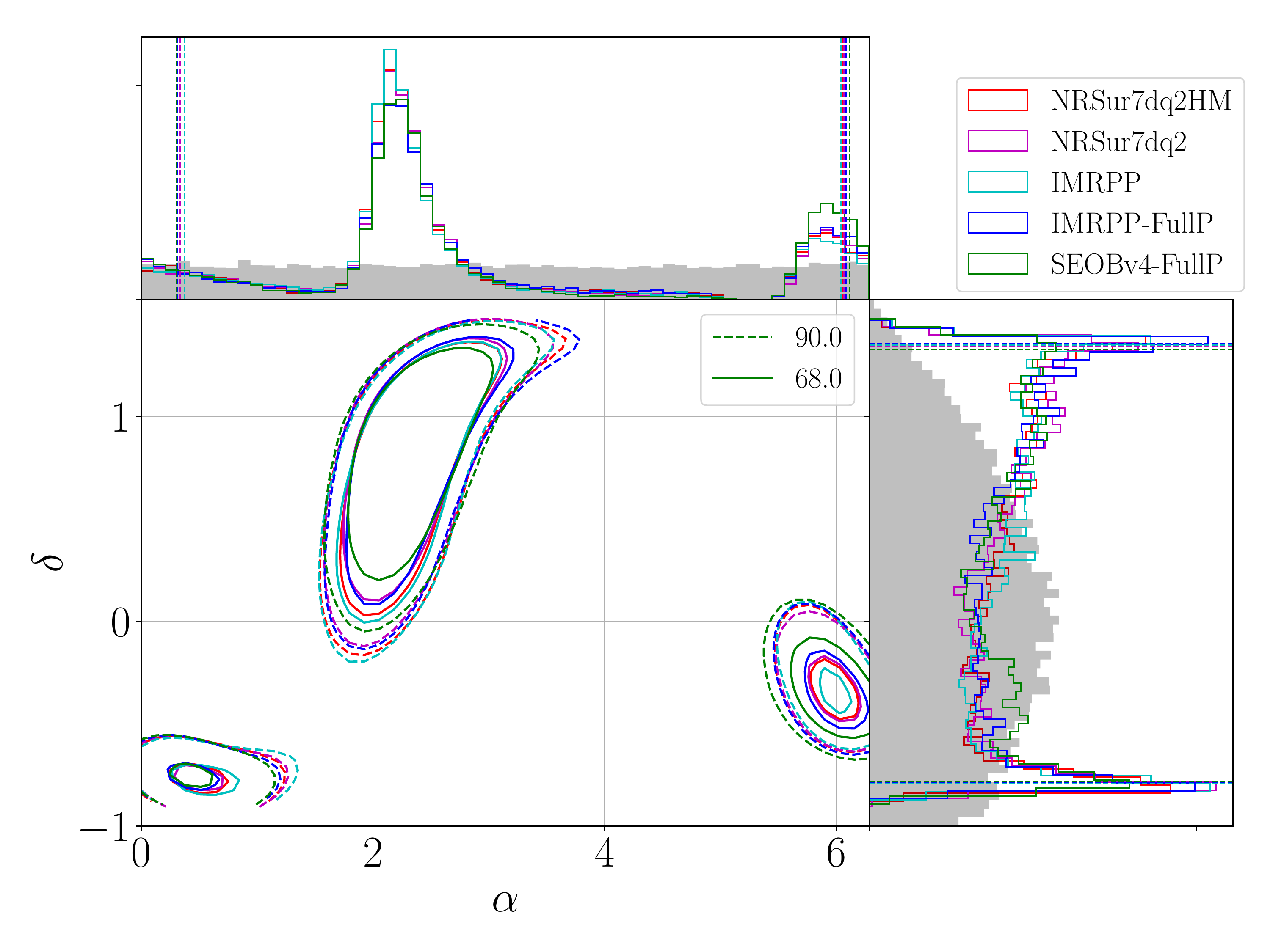}
 \caption{Estimated sky location for GW170104, using different approximants:
 \NRSurHM{}, \NRSurL{}, \Phenom{} (labeled IMRPP), and \SEOB{} (labeled SEOBv4).
 The X-FullP results correspond to an analysis with model X that allows for
 unrestricted mass ratios $1\leq q \leq 8$, and spin magnitudes up to
 $a_{1,2}\lesssim 0.89$ for \Phenom{} and $a_{1,2}\lesssim 0.98$ for
 \SEOB{}. For all others, we {\it a priori} restrict
 sampling to $1\leq q \leq 2$ and $0\leq a_{1,2}\leq 0.8$, i.e. to the range where NR
 surrogate models are valid.
 In all panels showing $1$-D posterior distributions, the shaded region shows our 
 prior belief. Vertical dashed lines in $1$-D posteriors mark $90\%$ credible regions.
 The 2-D posteriors show both the $90\%$ (dashed line)
 and $68\%$ (solid line) credible regions.
 \label{fig:ra_dec_68_90pc_NR_NRHM_PP_PPFullP_gw170104}}
\end{figure}
\begin{figure}
\centering
 \includegraphics[width=\columnwidth,clip=true,trim={8mm 5mm 0 0}]{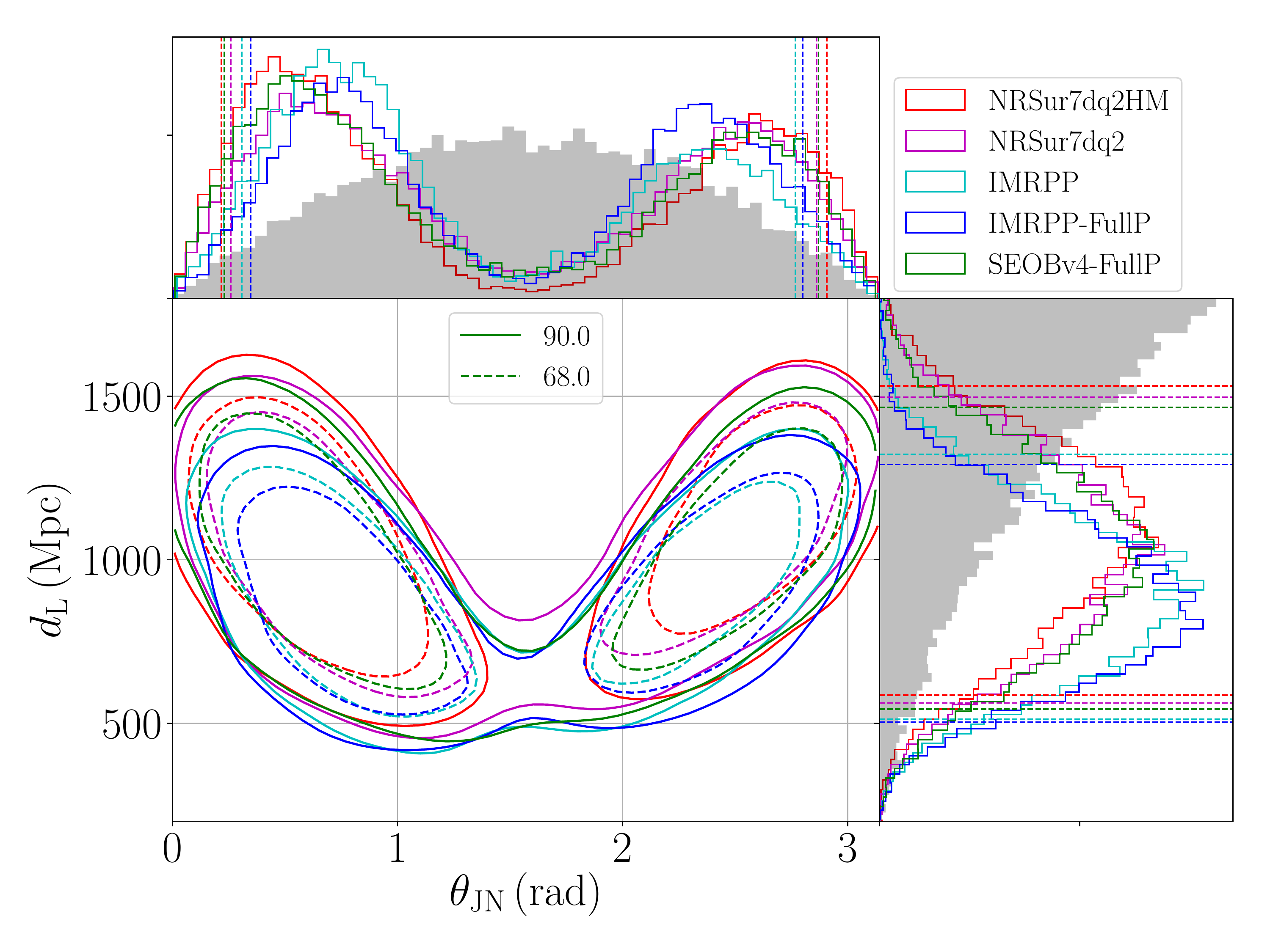}
 \caption{Estimated source orientation / luminosity distance
 for GW170104, using different approximants: \NRSurHM{}, \NRSurL{},
 \Phenom{} (labeled IMRPP), and \SEOB{} (labeled SEOBv4).
 All figure attributes are identical to
 Fig.~\ref{fig:ra_dec_68_90pc_NR_NRHM_PP_PPFullP_gw170104}.
 \label{fig:dist_incl_68_90pc_NR_NRHM_PP_PPFullP_gw170104}}
\end{figure}
\begin{figure*}
\centering
 \includegraphics[width=\columnwidth,clip=true,trim={4mm 5mm 0 0}]{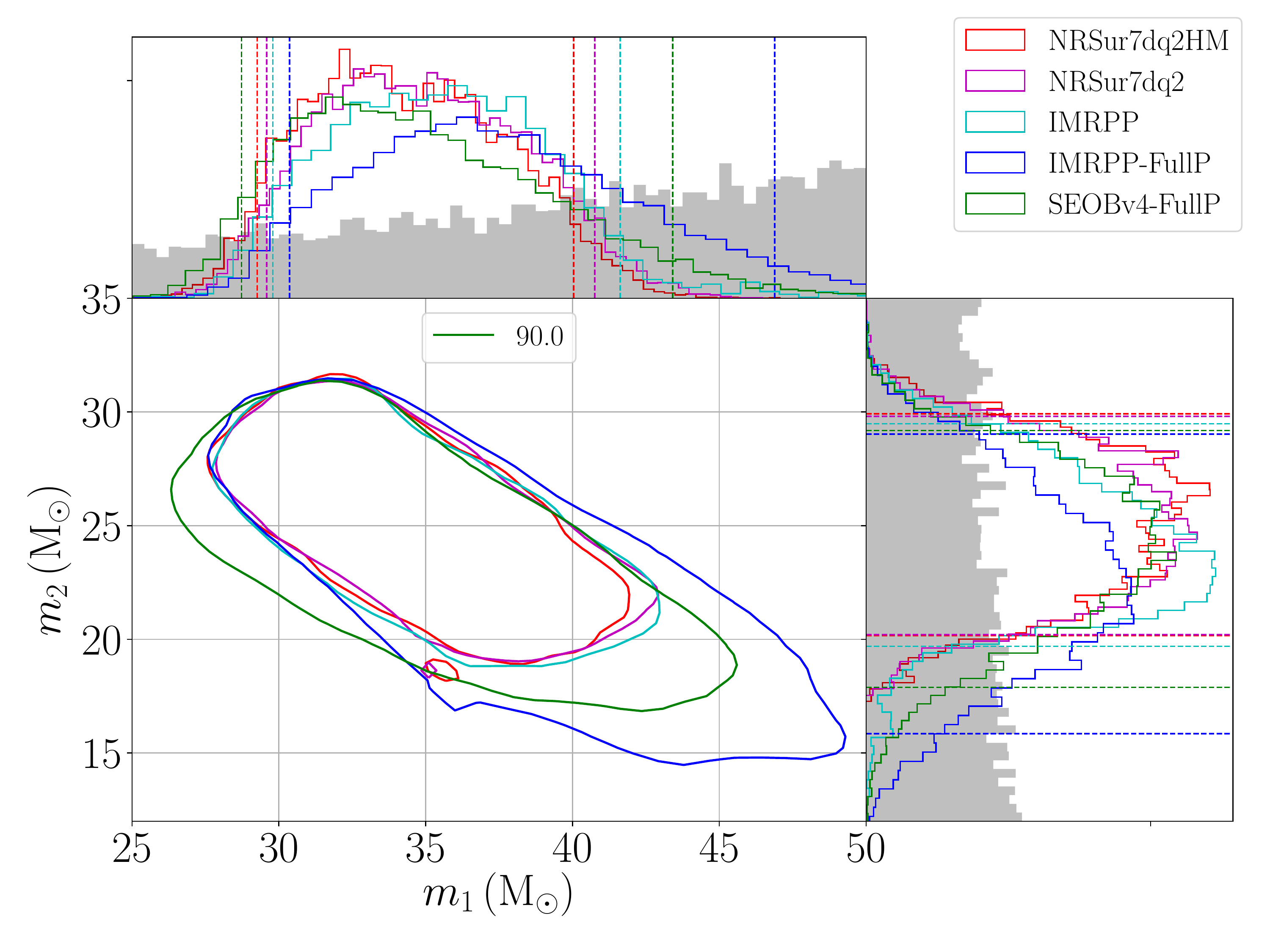}
 \includegraphics[width=\columnwidth,clip=true,trim={8mm 5mm 0 0}]{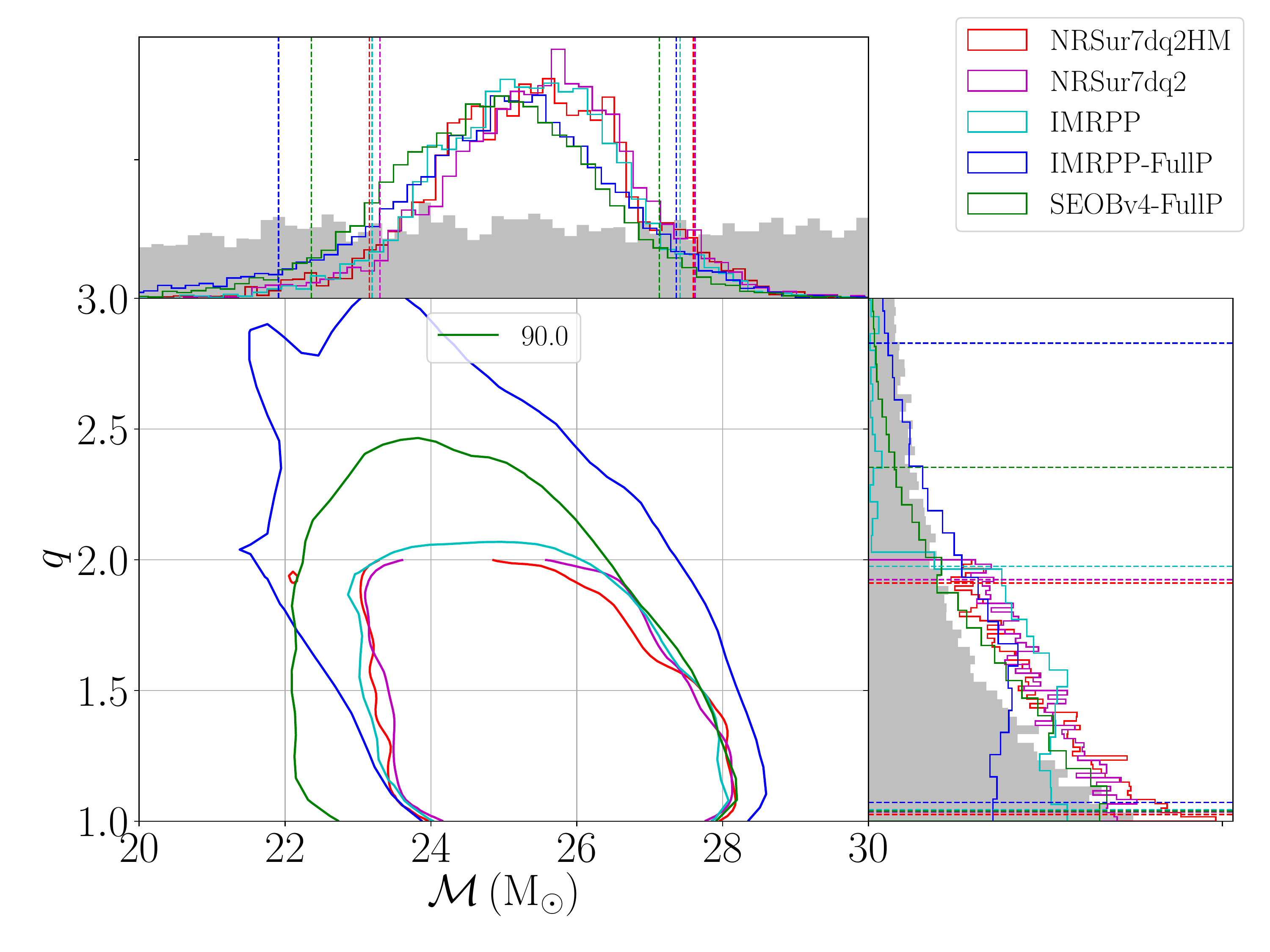}
 \caption{Estimated masses for GW170104, using different approximants:
 \NRSurHM{}, \NRSurL{}, \Phenom{} (labeled IMRPP) and \SEOB{} (labeled SEOBv4).
 The X-FullP results correspond to an analysis with model X that allows for
 unrestricted mass ratios $1\leq q \leq 8$, and spin magnitudes up to
 $a_{1,2}\lesssim 0.89$ for \Phenom{} and $a_{1,2}\lesssim 0.98$ for
 \SEOB{}. For all others, we {\it a priori} restrict sampling to $1\leq q \leq 2$
 and $0\leq a_{1,2}\leq 0.8$, i.e. to the range where NR surrogate models are
 valid. In all panels showing $1$D posterior distributions, the shaded region
 shows our prior belief. Vertical dashed lines in $1$D posteriors mark $90\%$
 credible regions. 
 The $2$D posteriors show $90\%$ credible regions as solid contours.
 \label{fig:m1_m2_90pc_NR_NRHM_PP_PPFullP_gw170104}}
\end{figure*}
\begin{figure}
\centering
 \includegraphics[width=0.4\textwidth,clip=true,trim={8mm 8mm 0 0}]{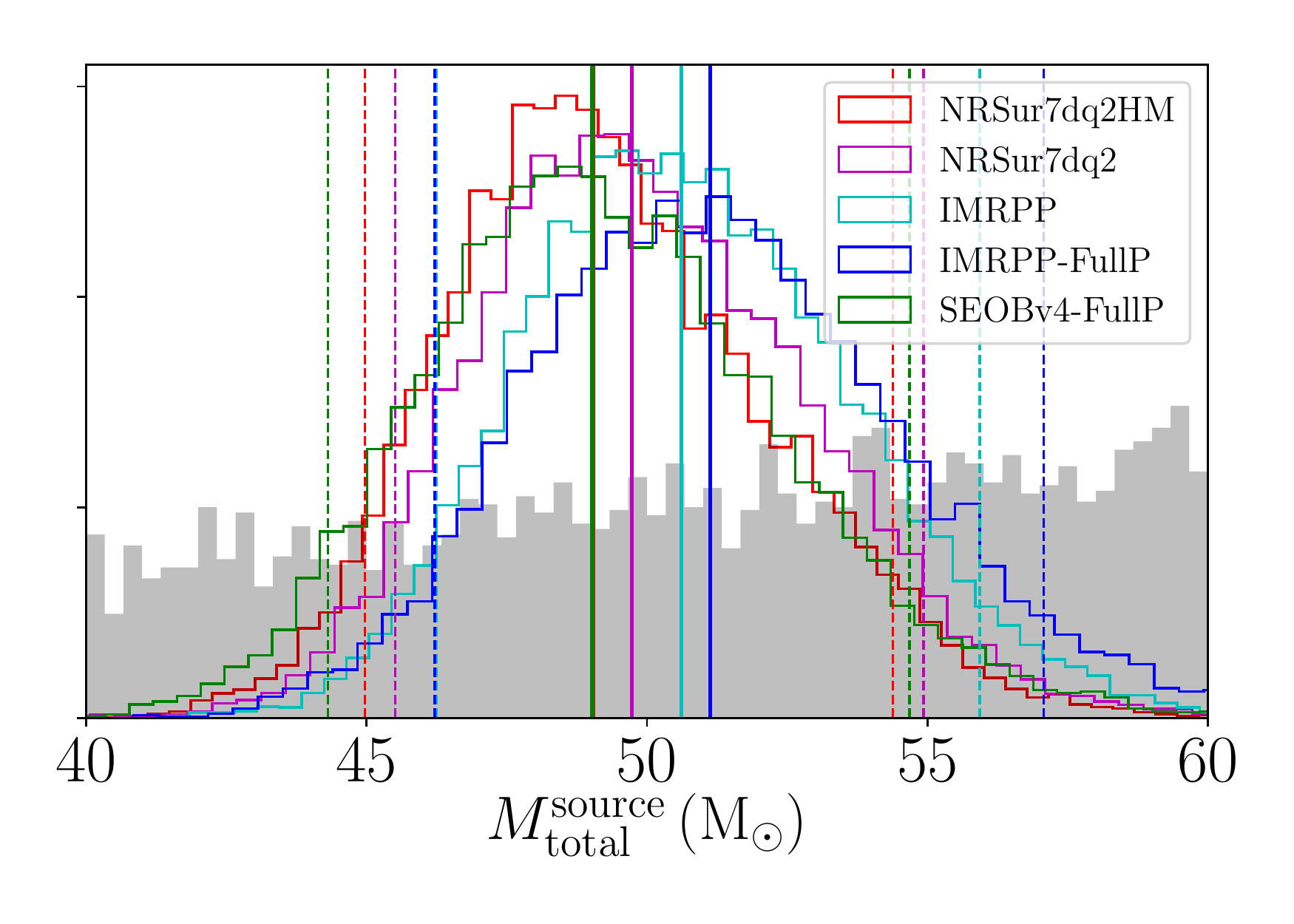}
 \caption{Total mass for GW170104, as measured in the source
 frame. Four different approximants are shown: \NRSurHM{},
 \NRSurL{}, \Phenom{} (labeled IMRPP), and \SEOB{} (labeled SEOBv4).
 All figure attributes are identical to
 Fig.~\ref{fig:m1_m2_90pc_NR_NRHM_PP_PPFullP_gw170104}.
 \label{fig:mtot_source_90pc_NR_NRHM_PP_PPFullP_gw170104}}
\end{figure}
\begin{figure*}
\centering
 \includegraphics[width=\columnwidth,clip=true,trim={8mm 5mm 0 0}]{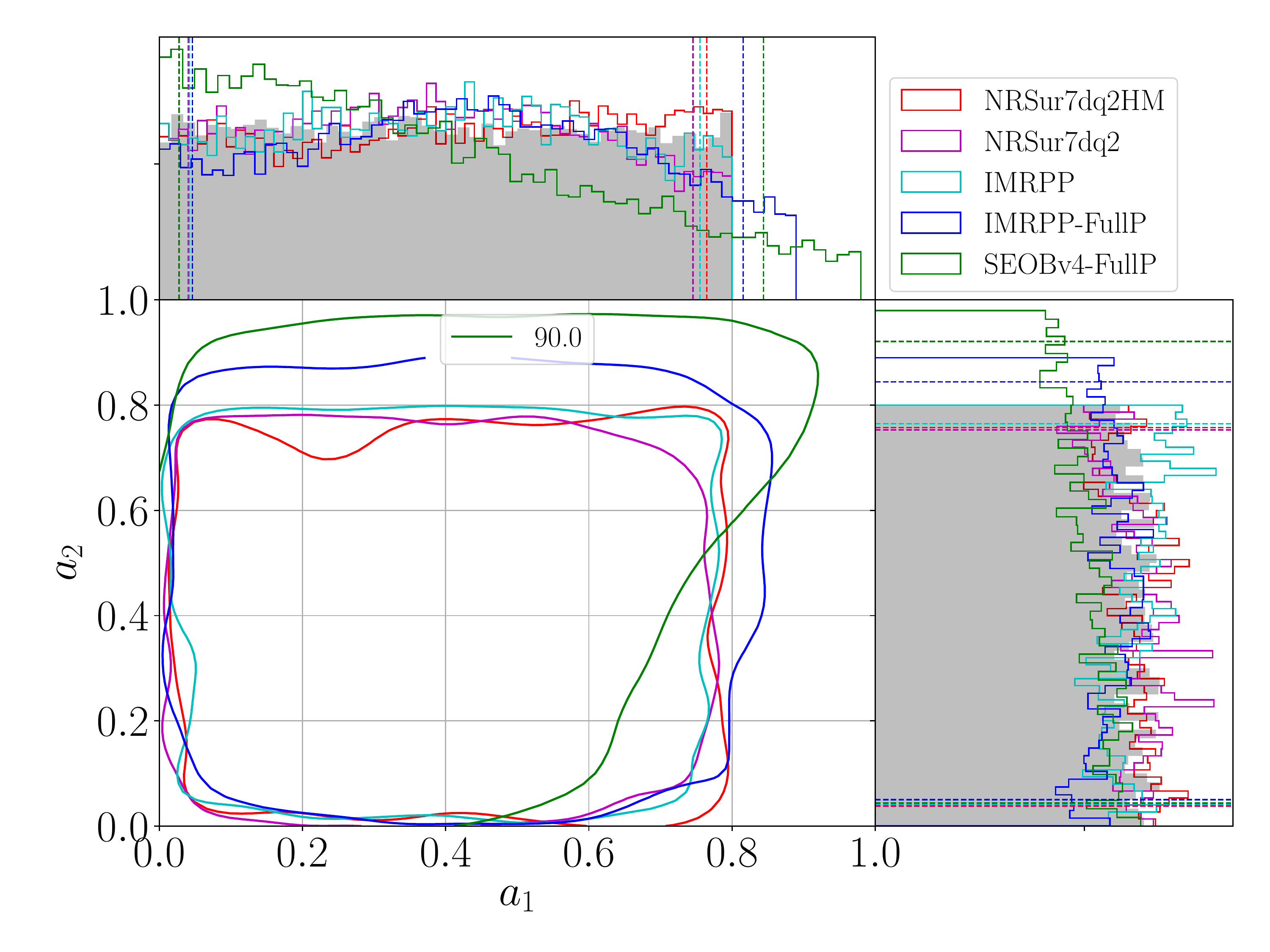}
 \includegraphics[width=\columnwidth,clip=true,trim={8mm 5mm 0 0}]{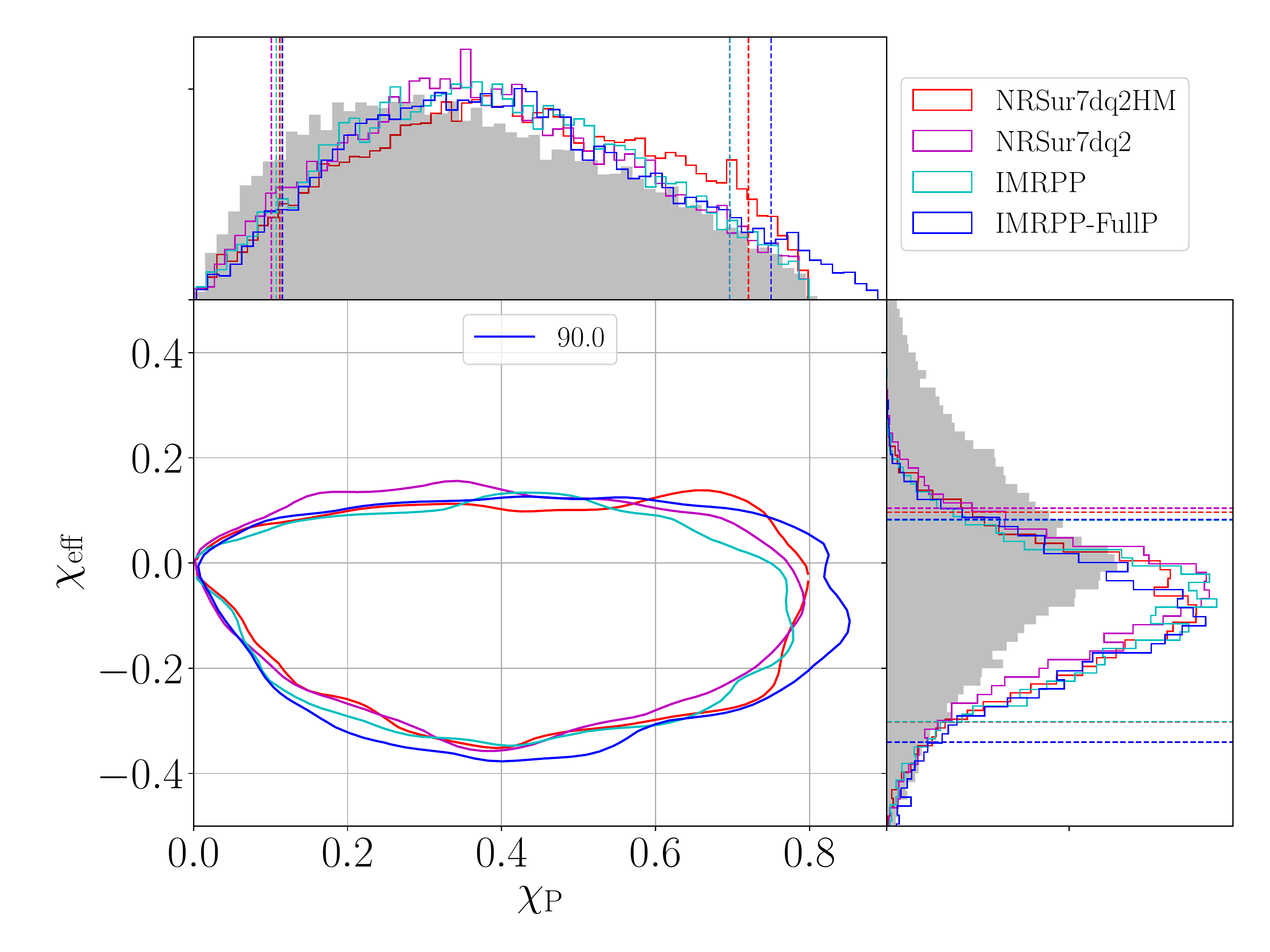}
 \caption{Estimated spins for GW170104, using different approximants and
 different prior probability distributions. Shown are spin magnitudes for
 both BHs and the tilt angles between BH spins and the orbital angular
 momentum at $f_\mathrm{ref}$. Figure attributes are identical to 
 Figs.~\ref{fig:ra_dec_68_90pc_NR_NRHM_PP_PPFullP_gw170104}
 and~\ref{fig:m1_m2_90pc_NR_NRHM_PP_PPFullP_gw170104}.
 \label{fig:a1_a2_tilt1_tilt2_90pc_NR_NRHM_PP_PPFullP_gw170104}}
\end{figure*}

\section{GW150914}\label{s1:gw150914}
Having assessed the performance of \NRSur{} surrogate models in characterizing
gravitational-wave signal sources in a fully Bayesian framework, we now analyze
the first ever observed GW event GW150914 with these models.
As before, we use the nested sampling algorithm in LALInference to perform
parameter recovery on the event, and use non-precessing \SEOB{} templates in
addition to \NRSur{} and \Phenom{}. We perform two analyses with
\Phenom{}: one where we artificially restrict sampling priors to the domain
of \NRSur{} and another where we do not. We, however, find that both analyses
furnish almost identical results, and therefore conclude that the effect of
sampling priors on GW150914's analyses is minimal. In the analysis with
\SEOB{}, we do not artificially restrict the sampling prior. As in the previous
section, we do not use the precessing EOB model of~\cite{Pan:2013rra} due to its
high computational cost.
Results from all of the above analyses are posterior probability distributions 
for physical parameters describing the GW source, which are shown in
Figs.~\ref{fig:ra_dec_68_90pc_NR_NRHM_PP_PPFullP_gw150914}~-~\ref{%
fig:a1_a2_tilt1_tilt2_90pc_NR_NRHM_PP_PPFullP_gw150914}.
In all of these figures, black curves in $1$D posterior distributions will show
prior distributions for respective parameters.

In Fig.~\ref{fig:ra_dec_68_90pc_NR_NRHM_PP_PPFullP_gw150914},
we show the recovery of the source's sky location angles (right ascension
$\alpha$ and declination $\delta$) by both \NRSur{} models and compare it with
those for semi-analytic models.
We immediately note that the recovery of sky location of GW150914's source
with \NRSurL{} is remarkably similar to that from semi-analytic models, but
adds little extra information.
In Fig.~\ref{fig:dist_incl_68_90pc_NR_NRHM_PP_PPFullP_gw150914},
we show the recovery of the source's
luminosity distance $d_L$ from LIGO detectors and its total angular momentum's
initial inclination $\tjn$ with respect to the line of sight. These two
parameters are strongly degenerate, as can be seen from the $2$-D posterior
slices showing $1$-$\sigma$ and $90\%$ credible regions for both.
Note the effect of higher-order waveform multipoles included in 
\NRSurHM{} on the measurement of both $d_L$ and $\tjn$. From their $90\%$
credible intervals in $1$D marginalized posteriors, we can see that
\NRSurHM{} places GW150914 at $\sim 530$Mpc, while other models, including
\NRSurL{}, place it at $\sim 430$Mpc. The primary LVC analyses of the event
also inferred the source to be at $\sim 410$Mpc~\cite{TheLIGOScientific:2016wfe}.
Therefore \NRSurHM{} locates the source of GW150914 about $25\%$ further away
than what other template models have so far.
The difference between $d_L$ posteriors estimated from \NRSurHM{} and \NRSurL{}
strongly implies that the difference in \NRSurHM{}'s luminosity distance
estimation is indeed due to the inclusion of higher-order waveform modes.
Similarly, the inclination angle is more precisely constrained by \NRSurHM{}
to be either face-on or face-off, with edge-on configurations being more
strongly disfavored by it than all other models. {\it These are some of the key
findings of this paper. They were inaccessible to the original LVC
analyses~\cite{TheLIGOScientific:2016wfe,Abbott:2016apu,Abbott:2017vtc}, which
were limited by modeling approximations and the availability of a sufficient
number of NR simulations.}
In Fig.~\ref{fig:dist_q_90pc_NR_NRHM_PP_PPFullP_gw150914} we show the
correlated posterior distribution for luminosity distance and mass ratio. We
see immediately that the increase in the estimated value of $d_L$ by \NRSurHM{}
is not an artifact of the model's restricted domain of validity since the region
of the posterior at large distances actually corresponds to nearly equal-mass
binaries.

Next, we show the recovery of mass parameters for GW150914 in
Fig.~\ref{fig:m1_m2_90pc_NR_NRHM_PP_PPFullP_gw150914}. While for
individual masses and mass ratio, \NRSurHM{} and \NRSurL{} give us very similar
posterior distributions to what we obtain from approximate waveform models,
the binary's chirp mass is estimated to be somewhat higher by both 
surrogate models. While this difference is marginal, it is consistent with
\NRSurHM{}'s estimation of luminosity distance to larger values, as the GW
signal strength depends on the ratio of the two. 
Further, given that the distance estimated by \NRSurHM{} and other waveform
models differs by $\sim 15\%$, we expect the measured source-frame mass to also
differ by $\delta M^\mathrm{source}\sim -M\delta z$ where $\delta z$ is the
corresponding difference in the inferred redshift of GW150914 (assuming standard
cosmology~\cite{Ade:2015xua}) between models, and $M$ is the estimated total mass.
This can be seen from Fig.~\ref{fig:mtot_source_90pc_NR_NRHM_PP_PPFullP_gw150914}
where we show the posterior distribution for the total mass of the binary in
its source frame. We find that \NRSurHM{}'s median estimate to be approximately
$0.5-1M_\odot$ {\it lower} than others, which is consistent with our estimate of
$-0.15\times 0.1\times 65\approx -1M_\odot$ (see also Table~\ref{tab:parameters}).

Finally, we focus on the recovery of binary spins for GW150914 in 
Fig.~\ref{fig:a1_a2_tilt1_tilt2_90pc_NR_NRHM_PP_PPFullP_gw150914}. The left
sub-figure shows marginalized $1$D and $2$D posteriors for individual BH spin
magnitudes (labeled $a_{1,2}\equiv |\vec{\chi}_{1,2}|$), and the right one
focuses on their effective-spin $\chieff:= (m_1\chi_{1z}+m_2\chi_{2z})/M$
and precessing-spin $\chi_p$~\cite{Schmidt:2014iyl} combinations. 
From the left sub-figure we note that spin recovery with \NRSurHM{} and \NRSurL{}
closely follows what we measure with our approximate precessing model
\Phenom{}. Restricted priors of \NRSur{} have no significant effect.
With \SEOB{} we find spin magnitudes to be constrained along the 
region with $|\vec{\chi}_{1}|\simeq|\vec{\chi}_{2}|$. This is as expected given
that the effective spin combination is constrained close to $0$ (right sub-figure),
which necessitates $\vec{\chi}_{1}\simeq -\vec{\chi}_{2}$ for a comparable-mass
binary such as this, and $\vec{\chi}_{1}$, $\vec{\chi}_{2}$ are always
(anti-)parallel for \SEOB{}.
From the right sub-figure, we immediately note that the surrogate model does
not recover any additional information about the binary's precessing spin
component, as its posterior appears to be sampling the prior with little
information being added by data. However, {\it it does constrain the source's
effective-spin to be somewhat closer to zero.} This is most clearly seen
by how the $1$D $90\%$ credible intervals differ between \NRSur{} and \Phenom{}.
As both \NRSurHM{} and \NRSurL{} provide for better estimation of $\chieff$,
we conclude that this may be because of additional spin information in the
surrogate models that is not included in the \Phenom{} model.

Overall, we conclude that {\it the use of} \NRSur{} {\it surrogate
improves the estimation of source distance and inclination for GW150914
substantially. This is primarily because of the inclusion of higher-order
GW modes. \NRSur{} also helps constrain GW150914's effective spin
somewhat better (closer to zero). This appears to be because \NRSur{}
models capture effects of BH spins on quadrupolar GW emission
better than approximate waveform models}.
These results are further quantified in Table~\ref{tab:parameters}, which can 
be directly compared with Table~I of~\cite{TheLIGOScientific:2016wfe}.

\section{GW170104}\label{s1:gw170104}

The second heavy binary black hole merger was detected by the two LIGO
detectors on January $4$, $2017$\footnote{The second actual detection
was GW151226~\cite{Abbott:2016nmj}. In the context of this paper, this
event was a ``light'' BBH merger, in contrast to ``heavy'' BBHs that we
focus on here.}. We perform identical analyses
on this event as we did for GW150914. Results are shown in
Figs.~\ref{fig:ra_dec_68_90pc_NR_NRHM_PP_PPFullP_gw170104}~-~\ref{fig:a1_a2_tilt1_tilt2_90pc_NR_NRHM_PP_PPFullP_gw170104}.

In Fig.~\ref{fig:ra_dec_68_90pc_NR_NRHM_PP_PPFullP_gw170104} we show the
recovery of GW170104's sky location with different models and choices of
priors. We immediately note that all models yield remarkably similar estimates
for its sky location, with the inclusion of higher-order modes in
\NRSurHM{} not yielding much additional information.
In Fig.~\ref{fig:dist_incl_68_90pc_NR_NRHM_PP_PPFullP_gw170104} we
show the recovery of source's luminosity distance $d_L$ and its initial
inclination angle $\tjn$ with respect to the line of
sight. Qualitatively similar to GW150914, we find that \NRSurHM{} narrows
the range of plausible $\tjn$ values to be closer to face-on and face-off
configurations (as opposed to edge-on ones). It estimates the source
to be located at a median distance of $1080$Mpc which is $20\%$ further
away than what we get when using approximate precessing and
non-precessing models models here ($882$Mpc) as well as in published
LVC analysis of the event ($880$Mpc)~\cite{Abbott:2017vtc}.
As was explicitly shown for GW150914 in
Fig.~\ref{fig:dist_q_90pc_NR_NRHM_PP_PPFullP_gw150914}, for GW1701014 too
we find that the increase in the estimated value of $d_L$ with \NRSurHM{} is
not due to the model's restricted domain of validity.

Next, we show the estimation of binary mass parameters for GW1701014 using
different models in Fig.~\ref{fig:m1_m2_90pc_NR_NRHM_PP_PPFullP_gw170104}.
We note that the $2$D posterior distribution for individual
BH masses has support at mass ratios larger than $q=2$ and therefore \NRSur{}
models only recover a fraction of the whole posterior. More specifically, it
appears that \NRSur{} models miss out on the low-$M$-high-$q$ portion of the
posterior. This would explain why even \NRSurL{} templates recover a slightly
higher value for $d_L$ than approximate models, as seen in the right panel of
Fig.~\ref{fig:dist_incl_68_90pc_NR_NRHM_PP_PPFullP_gw170104}, since an increase
in total mass estimate increases the estimated distance for a given signal
with fixed SNR.
However, the marginalized probability distributions estimated for chirp mass by
all precessing models are consistent, while the non-precessing \SEOB{} model
constrains it less stringently. Overall we find results from semi-analytic
models to be consistent with NR surrogate estimates.
Similar to GW150914, we expect the source-frame mass of GW170104 to be measured
differently by \NRSur{} than other waveform models.
We confirm this through Fig.~\ref{fig:mtot_source_90pc_NR_NRHM_PP_PPFullP_gw170104}
where we show the posterior distribution for the total mass of the binary in
its source frame. We find that \NRSurHM{}'s median estimate to be approximately
$1M_\odot$ lower than others, see Table~\ref{tab:parameters} for other mass
parameters measured in source frame.

In Fig.~\ref{fig:a1_a2_tilt1_tilt2_90pc_NR_NRHM_PP_PPFullP_gw170104} we
demonstrate how well we estimate component BH spins for GW170104 with the NR
surrogate model, and compare it with what we get from semi-analytic ones.
In the left sub-figure we show the estimation of spin magnitudes, while in the right
sub-figure we show the same for the effective spin $\chieff$ and precessing spin
$\chi_p$ combinations. While all models estimate $\chieff\simeq -0.1$ for 
this event, they recover little information for either $a_{1,2}$ or $\chi_p$,
with their respective $1$D posteriors following closely their sampling priors.
For all spin combinations considered, we note that the recovery from all models
is remarkably similar, despite the additionally restricted priors of \NRSur{}
models. For this signal, therefore, the \NRSur{} and approximate models
provide essentially identical spin information.

Overall, we conclude that with the NR surrogate model \NRSur{} we estimate the
source of GW170104 to be approximately $20\%$ further away than was previously
estimated using semi-analytic waveform models~\cite{Abbott:2017vtc}. The same
surrogate furnishes little extra information for source mass and spin parameters
of GW170104 though. Our results are summarized in Table~\ref{tab:parameters}.

%
\begin{table*}
\caption{Summary of parameters that characterize GW150914 and GW170104. For
model parameters we report the median value as well as the range of the
symmetric $90\%$ credible interval~\cite{Aasi:2013kqa}; where useful, we also
quote $90\%$ credible bounds. 
The source redshift and source-frame masses assume standard cosmology~\cite{
Ade:2015xua}. The spin-aligned \SEOB{} (labeled SEOB) and precessing
\Phenom{} (labeled IMRPP) waveform models are described in the text, as is
the NR surrogate labeled here NR$22$ (\NRSurL{}) and NRHM (\NRSurHM{}).
Results for the effective precession spin parameter $\chi_\mathrm{p}$
used by \Phenom{} are not shown as we effectively recover the prior;
see left panels of Figs.~\ref{fig:a1_a2_tilt1_tilt2_90pc_NR_NRHM_PP_PPFullP_gw150914}
and~\ref{fig:a1_a2_tilt1_tilt2_90pc_NR_NRHM_PP_PPFullP_gw170104}.
The \SEOB{}/\Phenom{} values stated here are directly comparable to Table~I
of~\cite{TheLIGOScientific:2016wfe} for GW150914 and to Table~I
of~\cite{Abbott:2017vtc} for GW170104, and are broadly consistent with
published LIGO analyses.
}
\begin{ruledtabular}
\begin{tabular}{l p{65mm} p{65mm}}
 & \hspace{8mm} GW150914 \newline SEOB / IMRPP / NR$22$ / NRHM & \hspace{8mm} GW170104 \newline SEOB / IMRPP / NR$22$ / NRHM \\
\hline
Detector-frame total mass $M/\Msun$ & $72.1^{+3.5}_{-4.0}$   /   $71.6^{+4.1}_{-3.8}$   /   $71.9^{+5.1}_{-3.1}$   /   $72.2^{+4.8}_{-3.1}$   &    $58.8^{+5.8}_{-5.4}$   /   $60.2^{+5.7}_{-5.3}$   /   $59.9^{+4.6}_{-5.3}$   /   $59.5^{+4.8}_{-4.6}$    \\
Detector-frame chirp mass $\mathcal{M}/\Msun$ & $31.2^{+1.5}_{-1.8}$   /   $30.9^{+1.9}_{-1.8}$   /   $31.0^{+2.4}_{-1.5}$   /   $31.2^{+2.0}_{-1.5}$   &   $24.9^{+2.2}_{-2.5}$   /   $25.1^{+2.3}_{-3.0}$   /   $25.5^{+2.1}_{-2.4}$   /   $25.4^{+2.2}_{-2.1}$  \\
Detector-frame primary mass $m_1/\Msun$ & $39.0^{+5.7}_{-3.5}$   /   $38.7^{+5.2}_{-3.5}$   /   $39.1^{+5.1}_{-3.3}$   /   $38.9^{+4.8}_{-3.2}$   &   $34.3^{+9.1}_{-5.7}$   /   $37.0^{+9.7}_{-6.6}$   /   $34.9^{+5.8}_{-5.3}$   /   $34.1^{+5.9}_{-4.8}$ \\
Detector-frame secondary mass $m_2/\Msun$ & $32.9^{+3.2}_{-4.8}$   /   $32.9^{+3.4}_{-5.1}$   /   $33.1^{+3.5}_{-5.0}$   /   $33.5^{+3.2}_{-5.2}$   &   $24.1^{+5.1}_{-6.1}$   /   $22.7^{+6.3}_{-6.6}$   /   $24.7^{+4.9}_{-4.7}$   /   $25.4^{+4.5}_{-5.1}$ \\
Detector-frame final mass $M_\mathrm{f}/\Msun$ & $68.7^{+3.1}_{-3.6}$   /   $68.2^{+3.7}_{-3.4}$   /   $68.5^{+4.5}_{-2.8}$   /   $68.8^{+4.2}_{-2.7}$   & $56.3^{+5.4}_{-4.9}$   /   $57.8^{+5.7}_{-4.9}$   /   $57.3^{+4.3}_{-4.8}$   /   $56.9^{+4.4}_{-4.2}$  \\
\rule{0pt}{4ex}%
Source-frame total mass $M^\mathrm{source}/\Msun$ & $66.2^{+3.9}_{-3.9}$   /   $65.6^{+4.1}_{-3.4}$   /   $65.5^{+4.6}_{-3.3}$   /   $64.9^{+3.8}_{-2.9}$   &  $49.0^{+5.7}_{-4.7}$   /   $51.1^{+5.9}_{-4.9}$   /   $49.7^{+5.2}_{-4.2}$   /   $49.0^{+5.2}_{-3.9}$ \\
Source-frame chirp mass $\mathcal{M}^\mathrm{source}/\Msun$ & $28.6^{+1.6}_{-1.7}$   /   $28.3^{+1.8}_{-1.5}$   /   $28.3^{+2.0}_{-1.5}$   /   $28.1^{+1.7}_{-1.4}$   &   $20.7^{+2.2}_{-2.0}$   /   $21.2^{+2.3}_{-2.3}$   /   $21.2^{+2.2}_{-1.9}$   /   $20.9^{+2.4}_{-1.7}$  \\
Source-frame primary mass $m_1^\mathrm{source}/\Msun$ & $35.8^{+5.4}_{-3.3}$   /   $35.5^{+5.0}_{-3.2}$   /   $35.6^{+4.6}_{-3.2}$   /   $35.0^{+4.5}_{-2.9}$   &   $28.6^{+8.3}_{-4.8}$   /   $31.5^{+9.0}_{-6.0}$   /   $29.0^{+5.6}_{-4.6}$   /   $28.3^{+4.8}_{-4.3}$ \\
Source-frame secondary mass $m_2^\mathrm{source}/\Msun$ & $30.2^{+3.2}_{-4.4}$   /   $30.2^{+3.1}_{-4.6}$   /   $30.1^{+3.3}_{-4.5}$   /   $30.1^{+3.0}_{-4.6}$   &   $20.0^{+4.4}_{-4.8}$   /   $19.2^{+5.4}_{-5.2}$   /   $20.6^{+4.4}_{-3.8}$   /   $20.9^{+4.3}_{-4.2}$    \\
Source-fame final mass $M_\mathrm{f}^\mathrm{source}/\Msun$ & $63.0^{+3.5}_{-3.5}$   /   $62.5^{+3.7}_{-3.0}$   /   $62.4^{+4.1}_{-3.0}$   /   $61.9^{+3.5}_{-2.7}$   &   $46.9^{+5.5}_{-4.4}$   /   $49.0^{+5.9}_{-4.6}$   /   $47.6^{+4.9}_{-3.9}$   /   $46.9^{+4.8}_{-3.7}$   \\
\rule{0pt}{4ex}%
Mass ratio $q$ & $1.2^{+0.4}_{-0.2}$   /   $1.2^{+0.4}_{-0.2}$   /   $1.2^{+0.4}_{-0.2}$   /   $1.2^{+0.4}_{-0.1}$   &  $1.4^{+0.9}_{-0.4}$   /   $1.6^{+1.1}_{-0.6}$   /   $1.4^{+0.5}_{-0.4}$   /   $1.3^{+0.6}_{-0.3}$   \\
\rule{0pt}{4ex}%
Effective inspiral spin parameter $\chi_\mathrm{eff}$ & $-0.01^{+0.11}_{-0.15}$   /   $-0.03^{+0.14}_{-0.16}$   /   $-0.02^{+0.15}_{-0.12}$   /   $-0.01^{+0.13}_{-0.12}$   &   $-0.12^{+0.21}_{-0.28}$   /   $-0.10^{+0.19}_{-0.23}$   /   $-0.07^{+0.17}_{-0.24}$   /   $-0.09^{+0.19}_{-0.20}$  \\
Dimensionless primary spin mag. $a_1$ & $0.40^{+0.43}_{-0.36}$   /   $0.26^{+0.50}_{-0.24}$   /   $0.27^{+0.43}_{-0.24}$   /   $0.27^{+0.43}_{-0.25}$   &  $0.32^{+0.53}_{-0.29}$   /   $0.42^{+0.40}_{-0.37}$   /   $0.38^{+0.36}_{-0.34}$   /   $0.41^{+0.35}_{-0.37}$  \\
Dimensionless secondary spin mag. $a_2$ & $0.50^{+0.44}_{-0.45}$   /   $0.30^{+0.48}_{-0.28}$   /   $0.29^{+0.44}_{-0.27}$   /   $0.30^{+0.44}_{-0.27}$   &   $0.45^{+0.47}_{-0.41}$   /   $0.46^{+0.39}_{-0.41}$   /   $0.38^{+0.38}_{-0.34}$   /   $0.39^{+0.37}_{-0.35}$   \\
Final spin $a_\mathrm{f}$ & $0.69^{+0.04}_{-0.07}$   /   $0.67^{+0.05}_{-0.06}$   /   $0.68^{+0.05}_{-0.05}$   /   $0.68^{+0.04}_{-0.05}$   &   $0.62^{+0.10}_{-0.15}$   /   $0.61^{+0.08}_{-0.18}$   /   $0.64^{+0.06}_{-0.10}$   /   $0.64^{+0.07}_{-0.09}$    \\
\rule{0pt}{4ex}%
Luminosity distance $D_\mathrm{L}/\mathrm{Mpc}$ & $423.4^{+178.0}_{-189.7}$   /   $434.1^{+149.2}_{-182.7}$   /   $472.0^{+167.5}_{-193.2}$   /   $538.5^{+140.2}_{-181.3}$   &  $1000.9^{+467.0}_{-457.9}$   /   $882.3^{+407.6}_{-376.9}$   /   $1016.4^{+469.8}_{-456.3}$   /   $1079.3^{+441.3}_{-487.1}$ \\
Source redshift $z$ & $0.09^{+0.03}_{-0.04}$   /   $0.09^{+0.03}_{-0.04}$   /   $0.10^{+0.03}_{-0.04}$   /   $0.11^{+0.03}_{-0.04}$   &   $0.20^{+0.08}_{-0.09}$   /   $0.18^{+0.07}_{-0.07}$   /   $0.20^{+0.08}_{-0.08}$   /   $0.21^{+0.07}_{-0.09}$  \\
\hline
Upper bound on primary spin mag. $a_1$ & $0.74$   /   $0.65$   /   $0.62$   /   $0.62$   &  $0.85$   /   $0.82$   /   $0.75$   /   $0.76$ \\
Upper bound on secondary spin mag. $a_2$ & $0.94$   /   $0.78$   /   $0.73$   /   $0.73$   &  $0.92$   /   $0.85$   /   $0.75$   /   $0.76$  \\
Upper bound on mass ratio $q$ & $1.56$   /   $1.55$   /   $1.53$   /   $1.52$   &  $2.34$   /   $2.79$   /   $1.93$   /   $1.90$ \\
\hline
\end{tabular}
\end{ruledtabular}
\label{tab:parameters}
\end{table*}

\section{Discussion}\label{s1:discussion}
%
The population
of binary black hole mergers that the LIGO-Virgo detector network has
observer
thus far comprises of many loud signals coming from ``heavy'' black hole
binaries, with each hole measuring around $20-30$ times the mass of the
Sun~\cite{LIGOVirgo2018:GWTC_1}.
Coalescing binaries of such heavy black holes radiate gravitational waves at
lower frequencies than their lighter counterparts, and therefore enter the
sensitive frequency band of current GW detectors only a few orbits before
they merge. During this pre-merger period of the two-body evolution that is 
visible to LIGO-Virgo, inspiraling binaries' orbits evolve rapidly from
being well approximated as a sequence of slowly shrinking spheres (or circles)
to being highly dynamical as both holes enter each other's strong-field regions
at highly relativistic velocities. Describing their motion in the pre-merger
regime, and consequently the form of emitted gravitational radiation, is beyond
the reach of traditional perturbative methods that typically rely on the
dynamical timescale of gravity being large and/or binary motion being
non-relativistic.

Fully numerical solutions of nonlinear Einstein equations is the most powerful
(and only) approach that can tackle the physics in the pre-merger regime. This,
however, comes at a non-trivial computational cost that precludes performing
numerical
simulations for an arbitrary number of binary mergers. With present day
technology and budgets, it is possible to perform approximately
$\mathcal{O}(10^3)$ simulations in a calendar year. However, when trying to
determine the physical parameters of the source of a BBH merger event from its
observed GW data, one typically needs to matched-filter the data against
$\mathcal{O}(10^{6-8})$ distinct GW templates\footnote{Recently proposed
grid-based methods~\cite{Lange:2017wki,Lange:2018pyp} can recover a {\it subset}
of binary parameters with much fewer ($\mathcal{O}(10^{3})$) template
evaluations. However, approximations used in these methods include 
interpolating Bayesian likelihood on unstructured high-dimensional
grids. The (physical and technical) impact of these approximations still
needs to be quantified more thoroughly.}. Therefore there is a
large gap between the demand of matched-filtering templates and their availability
through direct numerical simulations. This gap is traditionally bridged by
introducing {\it phenomenological} extensions to perturbative waveform models,
and {\it calibrating} these extensions to agree with numerical simulations 
where they can. Examples of such models would include those within the
Effective-one-body family~\cite{Buonanno99} and the phenomenological
family~\cite{Ajith-Babak-Chen-etal:2007}.
Although these models now span a fair region of the full $7$-dimensional
parameter space that describes arbitrary binary black hole coalescences, there
is always scope for inaccuracies in one corner or another~\cite{Kumar:2015tha,
Kumar:2016dhh}.
An ab-initio more accurate and reliable approach would be to develop a
$7$-dimensional numerical interpolant (a {\it surrogate} model) for the
gravitational-wave strain based on select numerical relativity simulations. Such
a surrogate model would have been too expensive to construct in the past.
With improvements to numerical relativity technology in recent years,
Blackman et al~\cite{Blackman:2017dfb,Blackman:2017pcm} developed the first
usable surrogate models based solely on numerical relativity simulations. They
span a restricted region of the full $7$-D binary black hole parameter space,
but within that region the model describes arbitrary precessing binary
orbits to NR-level accuracy.

The primary purpose of this paper is to use the numerical relativity-based
surrogate model \NRSur{}~\cite{Blackman:2017pcm} in a fully Bayesian framework
and demonstrate both its viability and efficacy in estimating source parameters
from heavy binary black hole merger signals. This work paves the way for future
surrogate models that will gradually span the entire parameter space for most GW
events. We also use the same surrogate~\cite{Blackman:2017pcm} to re-analyze
data from the first two
heavy BBH merger events: GW150914 and GW170104. While we find improvements in
the precision of measuring mass and spin parameters for these events' source
binaries, our primary finding is that both of these events were located
$15-20\%$ further away than what approximate waveform-model based analyses
have found, including published LIGO-Virgo
results~\cite{TheLIGOScientific:2016wfe,Abbott:2017vtc}.

We first perform controlled tests by injecting synthetic GW signals into
zero noise, reconstructing their parameters using \NRSur{} templates as
filters, and comparing the parameter recovery with both the true parameters
as well as what other waveform models furnish. We perform a total of
$48$ such injections that are described in Table~\ref{tab:injections005}.
The injected source parameters are varied as follows: mass ratio takes
on values $\in\{1.2, 1.5\}$, while the total mass is fixed to
$60M_\odot$; source spins are chosen from four precessing configurations;
source distance is varied over $\{500, 1000, 1500\}$Mpc, while source
inclination is allowed two values - one nearly face-on and the other
nearly edge-on; and sky location angles are chosen uniformly over a
$2$-sphere.
We use the full \NRSurHM{} to model synthetic signals, and use both
\NRSurL{} and \NRSurHM{}, in addition to \Phenom{} (a 
phenomenological model for spin-precessing binaries with an effective
description of spin degrees of freedom), as filter templates.
We use \Phenom{} in two configurations: first, where templates are
artificially restricted to be sampled with the same prior restrictions
for BBH parameters as numerical surrogates, and second, where they can
be sampled freely. 
We find that both total mass and mass ratio are better recovered by
both \NRSurL{} and \NRSurHM{} templates than by \Phenom{}. This is
noticeable in Fig.~\ref{fig:all_param_mean_errors_injections005}.
For BBH spins, we find that all models produce broadly consistent
results, with the effective spin being measured somewhat more
accurately by \NRSur{}. For all other intrinsic BBH parameters, including
the chirp mass, the \NRSur{} templates furnish results that are broadly
consistent with those from \Phenom{}.
Amongst extrinsic parameters,
we find that the degeneracy in measuring source distance and orbital
inclination is largely reduced by the addition of $\ell=\{3,4\}$ multipoles
in \NRSurHM{}, and with them we can recover both of these parameters
better than all other template models, including \NRSurL{}.
We find this improvement to be especially pronounced when the source
is highly inclined to the line of sight and therefore emits more strongly
in $\ell=\{3,4\}$ GW modes.
Past work applying higher-mode information from NR to GW parameter
estimation could only use it to measure a subset of source parameters,
and relied on the interpolation of Bayesian likelihood on unstructured
grids~\cite{Abbott:2016apu,Lange:2017wki,Lange:2018pyp}. We point out that
our tests described above are the first {\it comprehensive} usage
of higher-mode information from NR without additional approximations.
Overall, we observe that \NRSur{} templates improve estimation of BBH
parameters both with and without the inclusion of $\ell >2$ GW modes.
This is understandable since it models both the leading and sub-leading
order GW modes more accurately than approximate GW
models~\cite{Blackman:2017pcm}, while most approximate models do not
yet include $\ell>2$ GW modes.
Most of this improvement is moderate, however, and we expect it to be
more pronounced when the signals themselves have either a larger relative
contribution from $l\neq 2$ modes, such as for binaries with higher mass
ratios, or their sources have larger spin magnitudes\footnote{In the
latter case, however, the sampling priors imposed on spins can alter their
estimation appreciably~\cite{Chatziioannou:2018wqx} and  must be carefully
chosen.}. Our results therefore provide strong motivation to extend the
\NRSur{} model to span a larger range of binary mass ratios and black hole
spins.

From these controlled tests, we establish the viability of using NR
information directly in a traditional Bayesian parameter estimation 
framework for GW events. We next proceed to analyze the first BBH merger
event ever to be recorded: GW150914. We analyze it with our NR surrogates
- \NRSurHM{} and \NRSurL{} - in addition to \Phenom{} and \SEOB{}.
The latter two (or their variants) were used in the original published 
analyses for this binary~\cite{TheLIGOScientific:2016wfe}. We find that
with \NRSurHM{} we place the source of this event to be at a luminosity
distance of $\simeq 530$Mpc, which is about $25\%$ further away than
what other models estimate (including previous LVC analysis of the
event~\cite{TheLIGOScientific:2016wfe}). If we remove $\ell=\{3,4\}$ modes
and restrict the analysis to \NRSurL{}, we find that the measured
luminosity distance agrees with the originally estimated value, indicating
clearly that this new information is extracted
by the sub-dominant waveform multipoles in \NRSurHM{}. Simultaneously,
\NRSurHM{} also helps narrow down the allowed inclination configurations for
the source to be either face-on or face-off with more confidence than earlier.
Both of these improvements can be seen from
Fig.~\ref{fig:dist_incl_68_90pc_NR_NRHM_PP_PPFullP_gw150914}.
As would be consistent with a larger luminosity distance, the NR
surrogates estimate the chirp mass of GW150914 to be marginally higher
than what approximate model estimate. Consistent with a larger redshift,
\NRSurHM{} estimates GW150914's mass in its source frame to be
approximately $1\%$ lower than other models. Finally, with full GR
information implicitly contained within them, both \NRSurL{} and
\NRSurHM{} constrain the effective-spin of GW150914 more tightly
around $\chieff \simeq 0$ than previous estimates. Components of spin
that are orthogonal to the orbit and that contribute to its precession are not
constrained much better than phenomenological models, and this is as expected
because for short signals there is simply not enough time for the binary to
complete a few precession cycles. However, measurement of BH spins can be
sensitive to the choice of sampling priors employed~\cite{Chatziioannou:2018wqx,
Lange:2018pyp}. We defer an investigation of their effect on spin inferences
with \NRSurHM{} to future work.

Finally, we move on to the second heavy BBH merger event: GW170104. This event
differs from GW150914 in the sense that its measured posterior probability
distributions (by approximate models) for binary masses have support outside
the domain of validity of \NRSurHM{}, specifically for binaries with mass
ratios $q>2$. In practice, however, we find that this restriction does not
bias the recovery of other parameters by \NRSurHM{} in any noticeable manner.
Similar to GW150914, \NRSurHM{} constrains the luminosity distance to this
event to be approximately $20\%$ larger than what approximate models that
include only the dominant $\ell=|m|=2$ modes give~\cite{Abbott:2017vtc}.
The orientation of this source is constrained more tightly around face-on
or face-off configurations, with edge-on configurations being strongly
disfavored. The estimation of mass parameters is consistent between
\NRSur{} and other models, although the former can only recover a partial
posterior distribution for mass ratio. Lastly, we
note that the estimation of spin parameters by \NRSurHM{} remains remarkably
similar to what we get from \Phenom{} and \SEOB{}, and is therefore
consistent with them. The effective spin for the event is the only well-measured
spin parameter. It is constrained to be small but negative by all models.
For all other spin combinations, all models essentially recover
the sampling prior as the posterior, with data adding little information.

From our results it is clear that there are certainly advantages of using 
numerical relativity surrogates for following up heavy binary black hole 
merger events. One of them is the inclusion of accurate $\ell>2$ modes
in \NRSurHM{}, which facilitates the resolution of the luminosity distance -
inclination angle degeneracy. This degeneracy often leads to systematic
bias in providing point estimates of the distance to GW sources, which
subsequently percolates to the calculations of astrophysical binary merger
rates~\cite{Abbott:2016nhf}, estimation of Hubble's
constant~\cite{Nishizawa:2016ood}, etc. These applications could therefore
potentially benefit from \NRSurHM{}-based follow-ups of GW events.
Another benefit comes as improvement in the measurement of BBH masses by
\NRSur{} (both \NRSurHM{} and \NRSurL{}) as it captures even the
quadrupolar GW mode more accurately than approximate models. 
The primary disadvantage of using \NRSur{} templates for GW event follow-up
is its limited domain of applicability. One of the lines of active
research we are currently pursuing is to extend the domain of \NRSur{}
to model binary emitters with mass ratios $q>2$.
Finally, we note that the computational cost of using \NRSurHM{} in
parameter estimation is approximately $3-4\times$ the cost of using the 
frequency-domain \Phenom{} / reduced order model for \SEOB{}. A bulk
of this extra cost is due to the extra Fourier transform required to
transform each time-domain surrogate template to frequency-domain.
This extra cost can, however, be mitigated in two ways. The first is
to develop a reduced order model for the surrogate, along the lines
of~\cite{Purrer:2014fza}. Another is using a recently developed
{\it rapid} parameter estimation scheme~\cite{Lange:2018pyp} with
\NRSurHM{} templates.

Our results are encouraging and we propose for \NRSur{} and its follow-up
models to used in standard GW event follow-up analyses in order to maximize
the science output from GW detector data.
We provide full posterior samples (as supplemental materials)
from Bayesian parameter estimation of LIGO/Virgo data for GW150914 and GW170104,
with \NRSurHM{} and \Phenom{}, to enable further analysis by the community.
These can be obtained from
\url{https://github.com/prayush/GW150914_GW170104_NRSur7dq2_Posteriors}.

\begin{acknowledgments}
  We gratefully acknowledge support for this research at Cornell from the
  Sherman Fairchild Foundation and NSF grant PHY-1606654; at Caltech from
  the Sherman Fairchild Foundation and NSF grant PHY-1404569; at CITA from
  NSERC of Canada, the Ontario Early Researcher Awards Program, the Canada
  Research Chairs Program, and the Canadian Institute for Advanced Research;
  and at Princeton from NSF grant PHY-1305682 and the Simons Foundation.
  SEF was partially supported by NSF grant PHY-1806665.
  PK would like to thank FAPESP grant 2016/01343-7 for hospitality during his
  visits to ICTP-SAIFR where part of this work was completed.
  This research has made use of data, software and/or web tools obtained from
  the LIGO Open Science Center (https://losc.ligo.org), a service of LIGO
  Laboratory, the LIGO Scientific Collaboration and the Virgo Collaboration.
  LIGO is funded by the U.S. National Science Foundation. Virgo is funded by
  the French Centre National de Recherche Scientifique (CNRS), the Italian
  Istituto Nazionale della Fisica Nucleare (INFN) and the Dutch Nikhef, with
  contributions by Polish and Hungarian institutes.
\end{acknowledgments}


\FloatBarrier
\bibliography{References}

\end{document}